\documentclass[intlimits,twoside,a4paper]{article}

\usepackage[cp1251]{inputenc}
\usepackage{xcolor}
\usepackage[eqsecnum]{cmpj3}

\usepackage{bm}

\issue{2020}{23}{2}{23601}
\doinumber{10.5488/CMP.23.23601}

\title[Correlation functions in inhomogeneous mixtures]%
{Effects of fluctuations on correlation functions in inhomogeneous mixtures}
\author[A. Ciach, O. Patsahan, A. Meyra ]{A. Ciach\refaddr{label1},
        O. Patsahan\refaddr{label2},  A.  Meyra\refaddr{label3,label4}}
\addresses{
\addr{label1} Institute of Physical Chemistry,
Polish Academy of Sciences, Kasprzaka 44/52, 01-224 Warszawa, Poland
\addr{label2} Institute for Condensed Matter Physics of the National
Academy of Sciences of Ukraine,\\ 1 Svientsitskii St., 79011 Lviv,
Ukraine
\addr{label3} IFLYSIB (UNLP, CONICET), 59 No. 789, B1900BTE La Plata, Argentina 
\addr{label4}
  Depto de Ingenieria Mec\'{a}nica, UTN-FRLP, Av. 60 esq. 124, 1900 La Plata, Argentina
}
\date{Received January 14, 2020, in final form February 16, 2020}
\begin{document}

\maketitle

\begin{abstract}
Approximate expressions for correlation functions in binary inhomogeneous mixtures are derived in a framework
of the mesoscopic theory  [Ciach A., Mol. Phys., 2011, {\textbf{109}}, 1101]. Fluctuation contribution is taken into
account in a Brazovskii-type approximation. Explicit results are obtained for two model systems. In the two models,
the diameters of the hard cores of particles are equal, and  the interactions favour a periodic arrangement of 
alternating species A and B. However, the optimal distance between the species A and B is much 
different in the two models. Theoretical results for different temperature and volume fractions of the two 
components are compared with the results of Monte Carlo simulations, and the structure is illustrated by simulation
snapshots. Despite different interaction potentials and different length scale of the local ordering, properties of the 
correlation functions in the two models are very similar. 
\end{abstract}

\section{Introduction}

Biological and soft-matter systems are typically multicomponent and inhomogeneous. For different systems, the inhomogeneities in density or concentration may appear on different length scales, ranging from the scale set by the size of molecules
through a few- to a few tens or even hundreds of molecular diameters. For example, in ionic systems or in two-component mixtures of highly  charged colloid particles, the concentration difference between the positively and negatively charged ions or particles oscillates in space in the crystalline phases on the length scale set by the size of the ions or particles~\cite{blaaderen:05:0,Bartlett2005}. 
In ionic liquids (IL) or molten colloidal crystals, the inhomogeneities remain present, and are reflected in the oscillatory decay of the correlation function for the concentration difference (or charge density) on the same length scale ~\cite{Shimizu2015,Hayes2015,ciach:06:2,patsahan:07:0}.
In this case, the inhomogeneity or local order actually means that the distribution of the components in the majority of microscopic states in the disordered phase differs significantly from the random distribution. 

In the case of weakly  charged colloid particles with solvent-induced effective short-range attraction, clusters of particles of various sizes and shapes or other assemblies are formed, which leads to inhomogeneities on a mesoscopic length scale~\cite{stradner:04:0,campbell:05:0,Sweatman:14:0,candia:06:0,santos:17:0,litniewski:19:0}. Similar inhomogeneities or self-assembly into different aggregates can occur in other systems  with interactions between the particles of the form of the so-called `mermaid' or SALR potential~\cite{royall:18:0}, for example in globular proteins in water~\cite{stradner:04:0,bergman:19:0,falus:12:0}.  In the SALR class of potentials, the interactions consist of short range attraction (SA) and long-range repulsion (LR). At low temperature, a universal sequence of the ordered phases (lyotropic liquid crystals) was predicted ~\cite{ciach:08:1,ciach:13:0,edelmann:16:0,pini:17:0} and was found in simulations~\cite{zhuang:16:0} for increasing the volume fraction of the particles. When the density increases, there is observed the formation of a cluster crystal of cubic symmetry, hexagonal arrangement of cylindrical clusters, gyroid network of particles, lamellas, gyroid network of voids, cylindrical voids and spherical voids.
When the ordered phases melt, the aggregates start to move freely leading to position-independent density. 
The local order is reflected in the oscillatory decay of correlations on the length scale set by the size of the aggregates.
Importantly,  even though the average density is position independent, almost all  simulation snapshots show the formation of aggregates~\cite{Sweatman:14:0,litniewski:19:0,santos:17:0}. The random distribution of particles is not observed in practice. Only at high temperature, the aggregates become disintegrated, and the random structure domintes in the microscopic states of the disordered phase. In a fixed mesoscopic window of the macroscopic volume, the density is either much larger or much smaller than the average density, when an aggregate either enters or leaves the window. This means that very large fluctuations of the local density take place. 
As a result, the internal energy obtained by a proper averaging of the energy of the microscopic states differs significantly from the energy calculated in the mean-field (MF) approximation, i.e., for the average density. This is because when the aggregates of a size determined by the range of attraction are separated by distances larger than the range of repulsion, there are much more pairs of attracting particles and much less pairs of particles that repel each other than in the state with a position-independent density (homogeneously distributed particles). That is why the energy of the majority of states in the inhomogeneous disordered phase is much lower than the energy calculated for the disordered phase in MF. The latter is the same for homogeneous and inhomogeneous systems. The much overestimated internal energy leads to instability of the disordered phase with respect to density modulations, and to a continuous phase transition  in MF that in reality is absent. The spurious instability of the disordered inhomogeneous phase in MF leads to divergent correlation functions that in reality do not diverge. Mesoscopic fluctuations of the local density, representing a different density inside the clusters and between them, i.e., the variance of the  local density or concentration, should be taken into account to restore the stability of the disordered inhomogeneous phase and to lead correct correlation functions and the first-order phase transition to the ordered phases at lower temperature.

The effects of mesoscopic fluctuations can be taken into account within the field-theory developed by Brazovskii~\cite{brazovskii:75:0}. This theory was adapted to amphiphilic systems in~\cite{leibler:80:0,podneks:96:0}. A mesoscopic density-functional theory with the variance of the local density taken into account in the approach based on the Brazovskii theory was developed in~\cite{ciach:08:1,ciach:13:0,ciach:18:0} for one-component systems, and in~\cite{ciach:11:2} for mixtures. 
A similar approach was used for the study of ionic systems in ~\cite{ciach:06:2,patsahan:07:0,ciach:18:1}.

At the formal level, the variance of the local density can be obtained by solving the self-consistent equation for the inverse correlation function in the one-loop Hartree approximation~\cite{ciach:08:1,ciach:12:0,ciach:18:0}. The contribution to the grand potential associated with the presence of inhomogeneities is calculated at the same level of approximation. From the physical point of view, the variance of the local volume fraction describes the average deviation of the volume fraction inside the clusters or  between them from the volume fraction averaged over the whole volume. This average inhomogeneity leads to a negative contribution to the
internal energy, and to stabilization of the disordered phase
for a much larger part of the phase diagram than in MF. In a one-component system, this theory leads to a correct high-$T$ part of the phase diagram at a semi-quantitative level~\cite{ciach:18:0}. The mathematical form of the correlation function for the volume fraction in this theory is the same as in MF, namely
\begin{equation}
\label{Gr}
G(r)=A\re^{-\alpha_0 r}\sin(\alpha_1 r+\theta)/r,
\end{equation}
but the dependence of the parameters on thermodynamic state is much different: there is no spurious divergence of $A$ or $1/\alpha_0$. Equation~(\ref{Gr}) should describe the asymptotic decay of correlations when  the correlation function in Fourier representation, $\tilde G(k)$, takes a maximum for $k=k_0> 0$ even in the exact theory. It  can be obtained from the pole analysis of $\tilde G(k)$, by taking into account the pair of poles $\ri\alpha_0\pm \alpha_1$ with the smallest imaginary part. However, simulations show that equation~(\ref{Gr}) very well describes the correlation function for the volume fraction or for the charge density in the SALR or IL systems already for $r$ larger than one period of the damped oscillations~\cite{litniewski:19:0, otero:18:0}.

Much less attention was paid to inhomogeneous mixtures. In~\cite{ciach:11:2},  only the general formalism has been 
developed, and a few examples were considered only in MF. In this article we focus on the structure of the disordered
inhomogeneous phase  described by correlation functions for different components. In section~\ref{sec2}, we briefly summarize 
the Brazovskii-type formalism generalized to mixtures in~\cite{ciach:11:2}.  Next, we restrict ourselves to binary 
mixtures and derive approximate equations for the correlation functions in Fourier representation, $\tilde G_{\alpha\beta}(k)$. 
We restrict ourselves to equal sizes of the hard cores of the particles, and consider two particular examples of a binary mixture.
In the first example, section~\ref{sec3}, we assume interactions leading to inhomogeneities on a length scale of $\sim 10\sigma$,
where $\sigma$ is a molecular size.  We choose the `mermaid' potential between the like particles, but a pair of different 
particles interact with a `peacock' potential having an attractive tail and a repulsive head. The tails of the interactions 
correspond to screened electrostatic potentials (repulsion between the like-particles, and attraction between the opposite charges) and
the short-range part of the interactions favours a phase separation of the two species. 
`Two mermaids and a peacock' effective interactions can occur between oppositely charged hydrophilic and hydrophobic colloid particles suspended in a
near-critical mixture with ions~\cite{ciach:16:0}.

In section~\ref{sec4}, we assume that short-range 
interactions of the square-well are formed only between different species. This potential can lead to a gas-liquid phase transition,
as well as to the ordered phase resembling an ionic crystal, with the structure on the length scale of 
$\sigma$ \cite{ciach:11:2}.
We calculate three correlation functions in the disordered phase for several state points for two 
models  both above and below the MF instability. In order to verify the theoretical predictions and to visualize the structure for different state points, we have performed MC simulations for the models considered. We conclude  in section~\ref{sec5}.

\section{Formalism}\label{sec2}

The theory is based
on the mesoscopic formalism developed for inhomogeneous mixtures with $n$ components in~\cite{ciach:11:2}.
Instead of the number density $ \rho_{\alpha}$ of the $\alpha$-component,
we consider the volume fraction in the mesoscopic region around ${\bf r}$, 
$\zeta_{\alpha}({\bf r})$, and  deviations $\Delta \zeta_{\alpha}({\bf r})$ from the average value, 
$\bar\zeta_{\alpha}({\bf r})$. The volume fraction is 
more suitable in the mesoscopic theory, particularly for unequal sizes of the particles. 
When the ordering occurs on the length scale much larger than the size of the particles, 
then the volume fraction and the number density  of the particles with the volume $ v_{\alpha}$ are related by $\zeta_{\alpha}\approx \rho_{\alpha} v_{\alpha}$.

Fixed forms of the local volume fractions of all the components, 
$\{ \zeta\}=(\zeta_1,\ldots,\zeta_n)$, can be considered as a  constraint  imposed on the microscopic states. 
Grand potential in the presence of the above mesoscopic constraint  can be written in the form $\Omega_\text{co}=U-TS-\mu_{\alpha} N_{\alpha}$,
where $U,S, N_{\alpha}$ and $\mu_{\alpha}$ are the internal energy, entropy, the number of molecules and the chemical potential
of the species $\alpha$ respectively
in the system with the constraint of fixed  $\{ \zeta\}$ imposed on the microscopic densities.  $U$ is given by the  expression
\[
U[\{\zeta\}]=\frac{1}{2}\int_{\bf r_1}\int_{\bf r_2}
\bar V_{\alpha \beta}(|{\bf r}_1-{\bf r}_2|)\zeta_{\alpha}({\bf r}_1)\zeta_{\beta}({\bf r}_2), 
\]
where the  element $\bar V_{\alpha\beta}(r)$ of the matrix $  \bar{\bf V}(r)$  is the interaction potential 
between the species $\alpha,\beta$, properly rescaled when the volume fraction instead of the number density is considered. In addition, the interaction potentials are multiplied by the function that vanishes for $r=|{\bf r}_1-{\bf r}_2|$ smaller than the sum of the hard-cores radii, and is equal to 1 otherwise. This way we avoid contributions to the internal energy from the states with overlapping cores of the particles.
We further assume that  the entropy $S$ satisfies the relation $-TS=F_\text{h}$, where $F_\text{h}$ is the free-energy of the 
hard-core reference 
system. We assume here a local-density approximation. 

In this work we study the effects of fluctuations on the correlation functions ${\bf G}$, with the 
matrix elements $G_{\alpha\beta}({\bf r}_{\alpha},{\bf r}_{\beta})=
\langle\Delta \zeta_{\alpha}({\bf r}_{\alpha})\Delta \zeta_{\beta}({\bf r}_{\beta})\rangle $
representing the correlations between the volume fractions of the species
$\alpha$ and $\beta$  at the positions ${\bf r}_{\alpha}$ 
and ${\bf r}_{\beta}$, respectively. In order to include  fluctuations, 
we introduce the functional $\beta F[\{\zeta\}]$ of the form
\begin{equation}
\label{FF}
\beta F[\{\zeta\}]=\beta\Omega_\text{co}[\{\zeta\}]-
\log\left[ \int \text{D}\phi_1 \ldots \int \text{D}\phi_n \re^{-\beta H_\text{fluc}}\right] ,
\end{equation}
where $\phi_{\alpha}({\bf r})$ is the local fluctuation of the volume fraction of the component $\alpha$, and
\[
H_\text{fluc}[\{\zeta\},\{\phi\}]=\Omega_\text{co}[\{\zeta\}+\{\phi\}]-\Omega_\text{co}[\{\zeta\}].
\]
The functional (\ref{FF}) becomes equal to the grand potential,
when  $\{\zeta\}=\{\bar\zeta\}$, with $\{\bar\zeta\}=[\bar\zeta_{\alpha_1}({\bf r}),\ldots,\bar\zeta_{\alpha_n}({\bf r})]$
that satisfy the minimum condition for (\ref{FF}). By definition, $\langle \phi_{\alpha}\rangle=0$ 
when  $\{\zeta\}=\{\bar\zeta\}$.

Note that from (\ref{FF}) it follows that the vertex  functions (related to the direct correlation functions) defined by
\[
{\cal C}_{\alpha_1,\ldots,\alpha_j}({\bf r}_1,\ldots,{\bf r}_j)= 
\frac{\delta^j \beta F[\{\bar\zeta\}]}{\delta\bar\zeta_{\alpha_1}({\bf r}_1)\ldots\delta\bar\zeta_{\alpha_j}({\bf r}_j)}
\]
consist of two terms: the first one is the contribution from the fluctuations on the microscopic length scale
with frozen fluctuations on the mesoscopic length scale. This term is
\begin{equation}
\label{calCn0}
{\cal C}^\text{co}_{\alpha_1,\ldots,\alpha_j}({\bf r}_1,\ldots,{\bf r}_j)= 
\frac{\delta^j \beta \Omega_\text{co}[\{\bar\zeta\}]}{\delta\bar\zeta_{\alpha_1}({\bf r}_1)\ldots\delta\bar\zeta_{\alpha_j}({\bf r}_j)}.
\end{equation}
The second term is the contribution from the fluctuations on the mesoscopic length scale.

We focus on  the two-point inverse correlation function that satisfies the analog to the Ornstein-Zernicke equation,
\begin{equation}
\label{OZ}
{\bf C}={\bf G}^{-1}.
\end{equation}
In the lowest-order nontrivial approximation, the matrix elements of the inverse correlation function in Fourier representation
are given by~\cite{ciach:11:2},
\begin{eqnarray}
\label{C}
\tilde C_{\alpha\beta}(k)=\tilde C^\text{co}_{\alpha\beta}(k)+\frac{A_{\alpha\beta\gamma\delta}}{2}{\cal G}_{\gamma\delta}\,,
\end{eqnarray}
where  the first term is the function defined in (\ref{calCn0}) with $j=2$ in Fourier representation, 
summation convention is used in the whole article, and
%
\begin{eqnarray}
\label{calG}
{\cal G}_{\gamma\delta}=\int\frac{\rd{\bf k}}{(2\piup)^3}\tilde G_{\gamma\delta}(k).
\end{eqnarray}
In the case of the disordered phase, $\tilde C^\text{co}_{\alpha\beta}(k)
=\beta \tilde V_{\alpha\beta}(k)+A_{\alpha\beta}$.
The matrix $\tilde {\bf V}(k)$ with the elements $\tilde V_{\alpha\beta}(k)$ is 
the interaction potential between the species $\alpha,\beta$ in Fourier representation,
and
$A_{\alpha_1\ldots\alpha_j}$ is given by
\[
A_{\alpha_1\ldots\alpha_j}=\frac{\partial^j \beta f_\text{h}(\{ \zeta({\bf r})\})}
{\partial\zeta_{\alpha_1}({\bf r})\ldots\partial\zeta_{\alpha_j}({\bf r})}.
\]
Equations (\ref{OZ})--(\ref{calG}) should be solved self-consistently, which is not an easy task, especially for
periodic structures in multicomponent mixtures. 

In this work we focus on a disordered  inhomogeneous
phase in a binary mixture, with $\alpha,\beta=1,2$.
Inhomogeneities at the mesoscopic length scale indicate that the correlation functions 
in Fourier representation take a maximum for $0<k<2\piup$. We assume that the inhomogeneities 
occur on a well-defined length scale, and the peak of $\tilde G_{\gamma\delta}(k)$ is high and narrow.

Let us consider 
\begin{eqnarray}
\label{calG1}
{\cal G}_{\alpha\beta}=\int\frac{\rd{\bf k}}{(2\piup)^3}\frac{[ \tilde C_{\alpha\beta}(k)]}{\det \tilde {\bf C}(k)}\,,
\end{eqnarray}
where  $[ \tilde C_{\alpha\alpha}(k)]=\tilde C_{\beta\beta}(k)$ and
$[ \tilde C_{\alpha\beta}(k)]=-\tilde C_{\alpha\beta}(k)$ for $\alpha\ne\beta$. By symmetry, $\tilde C_{12}=\tilde C_{21}$.
For the considered functions with a high, narrow peak, the main contribution to the integral comes 
from the vicinity of the maximum.  
In general, $\tilde C_{\alpha\beta}(k)/\det \tilde {\bf C}(k)$
can take the maximum for different values of $k$ for different pairs of $\alpha,\beta$. 
Here, we focus on the case where the maximum of all the integrands
in (\ref{calG1}) is very close to the minimum  at $k=k_0$ of $\det \tilde {\bf C}(k)$,
and we can make the approximation
\[
{\cal G}_{\alpha\beta}=[ \tilde C_{\alpha\beta}(k_0)]{\cal G},
\]
where
\begin{eqnarray}
\label{calG3}
{\cal G}=\int\frac{\rd{\bf k}}{(2\piup)^3}\frac{1}{\det \tilde {\bf C}(k)}.
\end{eqnarray}

In the Brazovskii-type theory considered here, the $k$ dependence of $\tilde C_{\alpha\beta}$ comes only
from $\tilde V_{\alpha\beta}(k)$,
and we can write (\ref{C}) in the form
\begin{eqnarray}
\label{Cc}
\tilde {\bf C}(k)=\beta\tilde{\bf V}(k)+{\bf c}\,,
\end{eqnarray}
where the elements $ c_{\alpha\beta}$ of  ${\bf c}$ are 
\[
c_{\alpha\beta}=A_{\alpha\beta}+\frac{A_{\alpha\beta\gamma\delta}}{2}{\cal G}_{\gamma\delta}.
\]
We introduce the notation
\begin{eqnarray}
\label{Ccc}
\det \tilde {\bf C}(k)=\det {\bf c}+\beta \tilde W(k),
\end{eqnarray}
with
\begin{eqnarray}
\label{W}
\beta\tilde W(k)=\beta\Big[
\beta\det\tilde {\bf V}(k)+\tilde V_{11}(k)c_{22}+\tilde V_{22}(k)c_{11}-2\tilde V_{12}(k)c_{12}
\Big].
\end{eqnarray}

Close to a deep minimum, we can make the expansion 
\begin{eqnarray}
\label{detC}
\det \tilde {\bf C}(k)= D_0+\frac{\beta\tilde W''(k_0)}{2}(k-k_0)^2+\ldots\,,
\end{eqnarray}
where $k_0$ is determined by the equation
\begin{eqnarray}
\label{W'}
\beta \tilde W'(k_0)=0
\end{eqnarray}
and 
\begin{eqnarray}
\label{D0}
D_0
=\det\tilde {\bf C}(k_0).
\end{eqnarray}
From the approximation (\ref{detC}) and (\ref{calG3}), we obtain~\cite{ciach:11:2,ciach:12:0}
\begin{eqnarray}
\label{calGexpl}
{\cal G}\approx
\frac{k_0^2}{\piup\sqrt{2\beta \tilde W''(k_0)D_0}}.
\end{eqnarray}
In order to obtain an explicit equation for $k_0$, we take into account that from (\ref{Cc}) it follows that ${\bf c}=\tilde {\bf C}(k_0)-\beta\tilde {\bf V}(k_0)$.
From the above and (\ref{W'}), (\ref{W}) we obtain the equation
\begin{eqnarray}
\label{k0}
\tilde V_{11}'(k_0)\tilde C_{22}(k_0)+\tilde V_{22}'(k_0)\tilde C_{11}(k_0)
-2\tilde V_{12}'(k_0)\tilde C_{12}(k_0)=0
\end{eqnarray}
that contains the 3 unknowns $\tilde C_{\alpha\beta}(k_0)$. 
$D_0$ is expressed in terms of the above  unknowns [see (\ref{D0})], and finally
\begin{eqnarray}
\label{W''}
\tilde W''(k_0)=2\beta \det \tilde {\bf V}'(k_0)+\tilde V_{11}''(k_0)\tilde C_{22}(k_0)+
\tilde V_{22}''(k_0)\tilde C_{11}(k_0)-2\tilde V_{12}''(k_0)\tilde C_{12}(k_0).
\end{eqnarray}

In order to determine ${\bf C}$ in this approximation from (\ref{Cc}), we need 3 equations
in addition to equation~(\ref{k0}), because we have 4 unknowns, $k_0$ and $\tilde C_{\alpha\beta}(k_0)$. 
Using (\ref{C}), (\ref{calGexpl}), (\ref{D0}) and (\ref{W''}) we obtain 3 equations 
with the unknowns $k_0$ and  $\tilde C_{\alpha\beta}(k_0)$,
\begin{eqnarray}
\label{C(k0)}
\tilde C_{\alpha\beta}(k_0)= \beta\tilde V_{\alpha\beta}(k_0)+ A_{\alpha\beta}+
\frac{k_0^2A_{\alpha\beta\gamma\delta}[\tilde C_{\gamma\delta}(k_0)]}{2\piup\sqrt{2\beta \tilde W''(k_0)D_0}}.
\end{eqnarray}
Equations~(\ref{k0}) and (\ref{C(k0)}) form a closed set of 4 equations for 4 unknowns, when the expressions 
(\ref{D0}) and (\ref{W''}) for $D_0$ and $\tilde W''$ are used, and $[\tilde C_{\gamma\delta}(k_0)]$ are defined below (\ref{calG1}).

Once $\tilde C_{\alpha\beta}(k_0)$ are determined, the correlation functions can be obtained from the equation
\begin{eqnarray}
\label{C(k)}
\tilde C_{\alpha\beta}(k)=  \tilde C_{\alpha\beta}(k_0) +\beta[\tilde V_{\alpha\beta}(k)-\tilde V_{\alpha\beta}(k_0)].
\end{eqnarray}

The approximate expression (\ref{calGexpl}) is valid only when the correlation functions in Fourier representation have a pronounced maximum at $k=k_0$. In such a case, $\tilde V_{\alpha\beta}(k)-\tilde V_{\alpha\beta}(k_0)$ in (\ref{C(k)}) can be expanded about $k_0$, and the expansion can be truncated. We should stress that the theory is not valid at high temperature, where no inhomogeneities at a well defined length scale are present.

In the next two sections we consider a binary mixture and
assume the same size of the spherical hard cores of the particles of the two kinds.
For the free energy of the hard-sphere reference system we assume $F_\text{h}=\int \rd{\bf r} f_\text{h}[\zeta_1({\bf r}),\zeta_2({\bf r})]$ with the Carnahan-Starling approximation for the free energy density $f_\text{h}(\zeta_1,\zeta_2)=\zeta_1\ln \zeta_1+\zeta_2\ln \zeta_2 + 
f_\text{ex}(\zeta)$, 
where $\zeta=\zeta_1+\zeta_2$ and
\[
f_{\text{ex}}(\zeta)=
\frac{6\zeta}{\piup}\left[ 
\frac{4\zeta-3\zeta^2}{(1-\zeta)^2}-1
\right] .
\]
For equal sizes, we have   $A_{12}=A_2(\zeta)$, $A_{1112}=A_{1122}=A_{1222}=A_4(\zeta)$, and $A_{\alpha\alpha}=A_2(\zeta)+1/\zeta_{\alpha}$, $A_{\alpha\alpha\alpha\alpha}=A_4(\zeta)+2/\zeta_{\alpha}^3$, where $A_n(\zeta)=\rd^n f_{\text{ex}}(\zeta)/\rd^n\zeta$. We use the diameter of the particles as the length unit.

We assume different interaction potentials that can lead to different inhomogeneities on different length scales.
\section{The case of interaction  potentials  $V_{11}=V_{22}=-V_{12}=V$}\label{sec3}

In this section we assume interaction potentials that  are a simplified version of the interactions between charged
colloid particles of equal sizes but of different sign of the charge, and different chemistry of the two species. 
For example, the species 1 can be `hydrophilic', while the species 2 `hydrophobic'. 
 The like-particles attract each other at short distances  due to a similar chemistry, but at large distances they repel
each other because of the screened electrostatic repulsion between like charges. Particles of a different kind
attract each other at large distances due to the screened electrostatic potential between opposite charges. 
We assume the short-range repulsion to enhance the tendency for  demixing  of uncharged particles. This kind of
interactions between the like-particles is known as a `mermaid' or SALR potential. Note that $V_{12}$ is repulsive at
short distances and attractive at large distances, i.e., it has a repulsive head and a attractive tail (like a peacock). 
In experiment, the  `two mermaids and a peacock' effective interactions can be obtained when 
the oppositely charged hydrophilic and hydrophobic colloid particles are immersed in a near-critical mixture of water and oil,
for example lutidine. Critical concentration fluctuations lead to the Casimir potential between confining 
surfaces~\cite{ciach:16:0}.
This potential is attractive for the like-surfaces, 
and repulsive between the hydrophilic and the hydrophobic surface. 
The range of the Casimir potential, equal to the bulk correlation length, $\xi_\text{b}$, can be tuned by temperature.
When  $\xi_\text{b}$ is smaller than the Debye length of the screened electrostatic potential, 
and the charge of the particles is properly chosen,
the `two mermaid and a peacock' potential can be created. 

\subsection{Theory}
To simplify the calculations, we assume that
the interactions between the like-particles are the same in this model,  $V_{11}=V_{22}=V$, 
and for a different kind of particles we assume $V_{12}=V_{21}=-V$. 
In our mesoscopic theory $V(r)=0$ for $r<1$, and for $r>1$ we assume the double-Yukava potential with 
short-range attraction and long-range repulsion, 
\begin{equation}
\label{int_pot_r}
V(r)=-\frac{K_1}{r}\re^{-\kappa_1r}+\frac{K_2}{r}\re^{-\kappa_2r}.
\end{equation}
Here, we do not try to model any particular system. Our aim is to verify the predictions of the mesoscopic
theory for the case of inhomogeneities present at the length scale larger than the size of the particles. 
We choose  $K_1=1,K_2=0.2,\kappa_1=1,\kappa_2=0.5$ that leads to relatively large clusters, with diameter $\sim 5\sigma$.
The amplitude of the attractive part of $V$ sets the energy unit, and we use the dimensionless temperature $T^*=k_\text{B}T/K_1$.
The high symmetry of the interaction potentials greately simplifies the calculations.

In MF, after some algebra we obtain the following expressions for the correlation functions  in Fourier representation (proportional to structure factors)
\begin{align*}
\tilde G_{11}^\text{co}&=G_0+\frac{(\zeta+c)^2[1+2A_2(\zeta-c)]^2}{D^\text{co}(k)}\,,
\nonumber \\
\tilde G_{22}^\text{co}&=G_0+\frac{(\zeta-c)^2[1+2A_2(\zeta+c)]^2}{D^\text{co}(k)}\,,
\nonumber \\
\tilde G_{12}^\text{co}&=G_0-\frac{(\zeta^2-c^2)\left[ (1+A_2\zeta)^2-A_2^2c^2\right] }{D^\text{co}(k)}\,,
\end{align*}
where the constant term leading to the Dirac delta function in real space is
\[
G_0=\frac{\zeta(1-c^2/\zeta^2)}{4[1+A_2\zeta(1-c^2/\zeta^2)]}\,,
\]
\begin{equation}
D^\text{co}(k)=4(1+A_2\zeta)^2\zeta
\left[ 
1-\frac{A_2c^2}{\zeta(1+A_2\zeta)}
\right] 
\left\lbrace   1+\zeta\left[ 
1-\frac{A_2c^2}{\zeta(1+A_2\zeta)}
\right]\right\rbrace,
\end{equation}
$\tilde V(k)$ is the interaction potential (\ref{int_pot_r}) in Fourier representation with the explicit expression given for example in~\cite{ciach:10:1}, and we have introduced $c=\zeta_1-\zeta_2$.
\begin{figure}[!b]
	\includegraphics[scale =0.35]{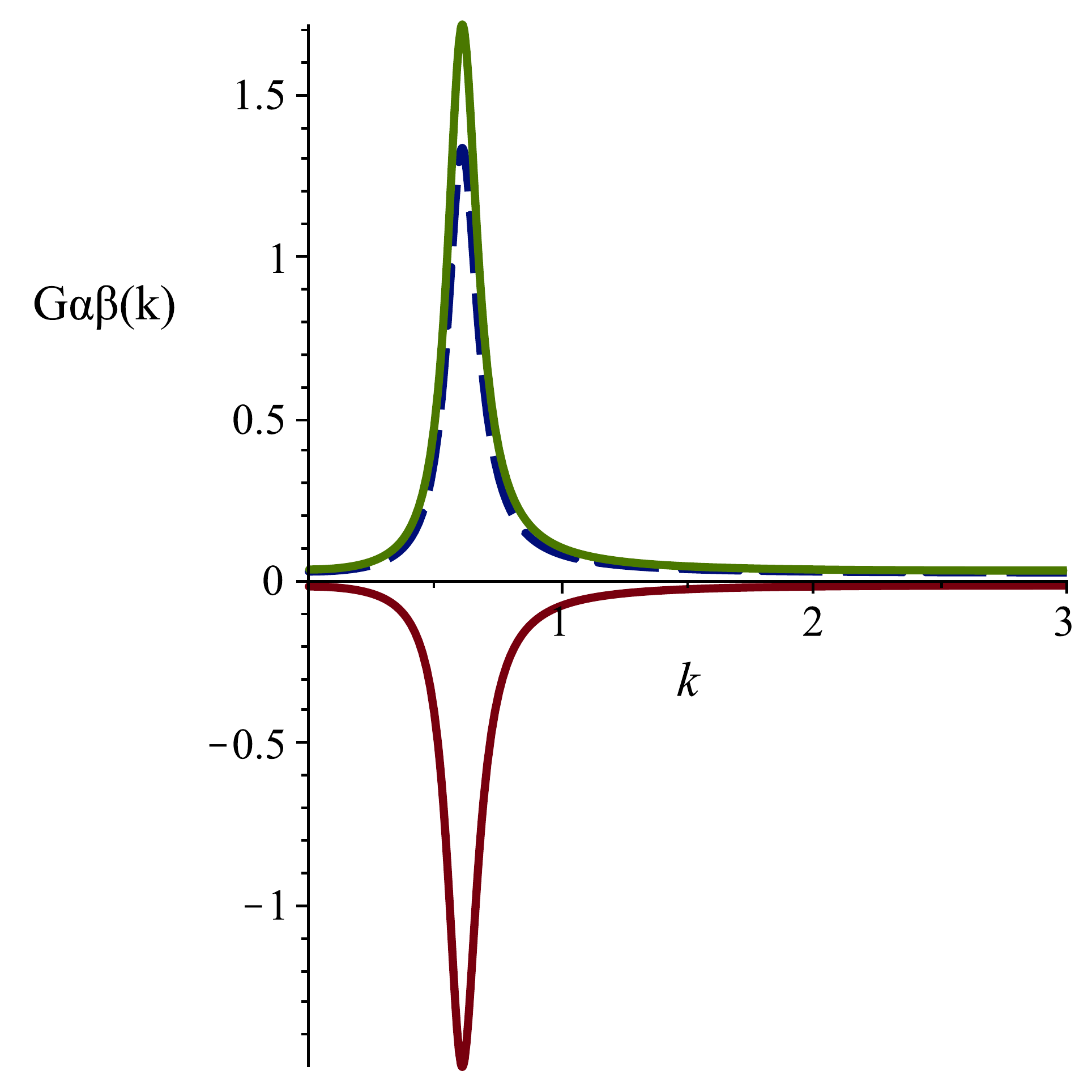}
	\includegraphics[scale =0.35]{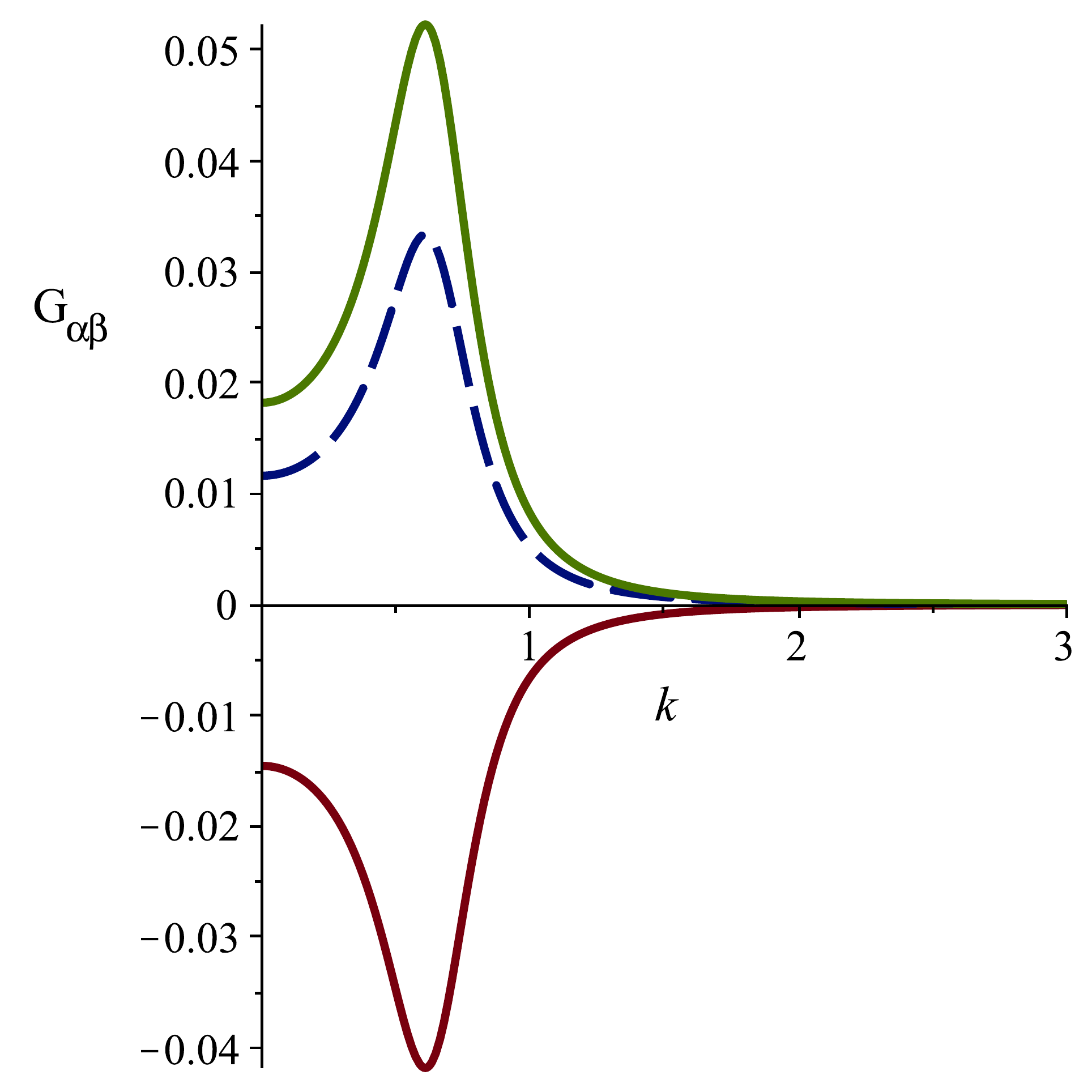}
	\caption{(Colour online) Correlation functions  in Fourier representation  for $T^*=0.28$, $\zeta=0.1$ and $c=0.02$ in MF (left-hand panel) and with the effect of fluctuations taken into account (right-hand panel). Upper solid lines: $\tilde G_{11}(k)$,   dashed lines: $\tilde G_{22}(k)$, and  lower solid lines: $\tilde G_{12}(k)$.
	}
	\label{fig_G(k)}
\end{figure}

\begin{figure}[!t]
	\includegraphics[scale =0.3]{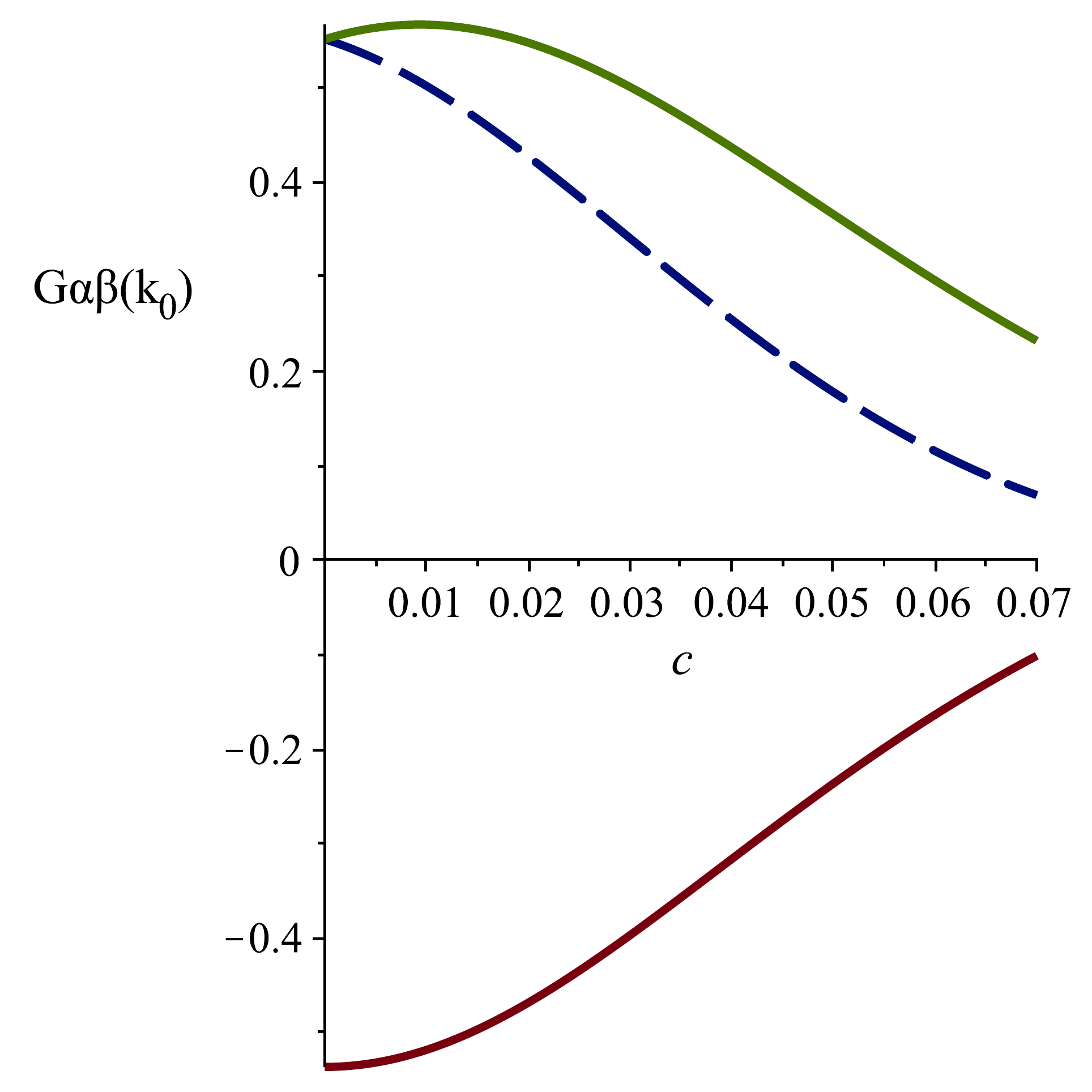}
	\includegraphics[scale =0.3]{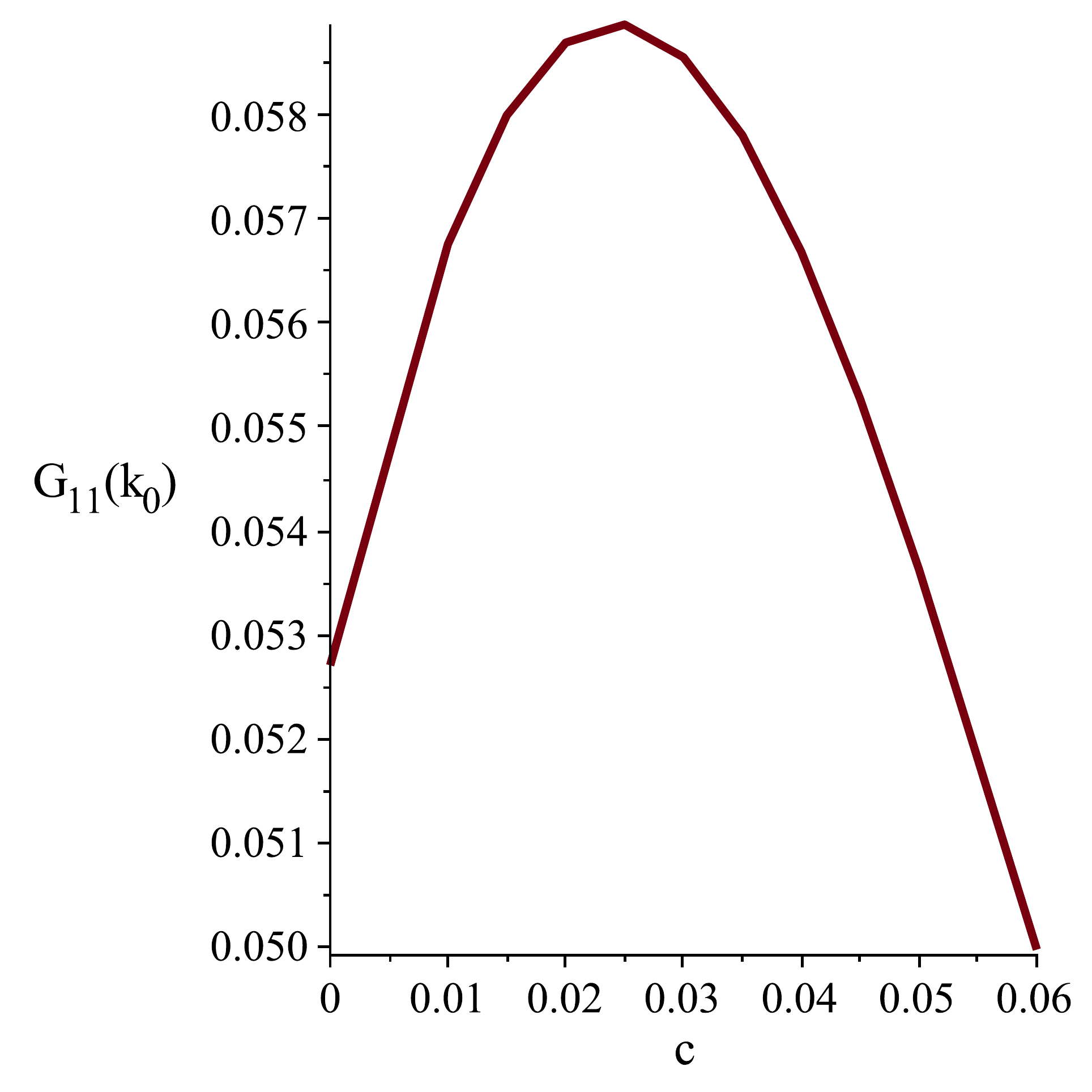}
	\caption{(Colour online) Left-hand panel: $\tilde G_{11}(k_0)$ (upper solid line), $\tilde G_{22}(k_0)$ (dashed line) and $\tilde G_{12}(k_0)$ (lower solid line)  in MF.
		Right-hand panel:  $\tilde G_{11}(k_0)$ with the effect of fluctuations taken into account.   $\tilde G_{22}(k_0)$ and $\tilde G_{12}(k_0)$ are monotonous functions of $c$. 
		$T^*=0.28$, $\zeta=0.1$ and $c$ is the difference in the volume fraction of the species 1 and 2.
	}
	\label{fig_G(k0)}
\end{figure}

In MF, the disordered phase becomes unstable with respect to oscillatory modulations of the volume fractions at the $\lambda$-surface given by 
\[
T^*(\zeta,c)=-\tilde V(k_0)\zeta\left\lbrace  
1-\frac{A_2(\zeta)c^2}{\zeta[1+A_2(\zeta)\zeta]}
\right\rbrace  ,
\]
therefore, we restrict ourselves to $T^*$ above the $\lambda$-surface, i.e., to the stability region of the disordered phase. 
Because of the symmetry of interactions, we assume that the majority component is the species 1, and consider only $c>0$.
The correlation functions, $\tilde G_{\alpha\beta}^\text{co}(k)$, are shown in figure~\ref{fig_G(k)} for $T^*=0.28$,
total volume fraction of the particles $\zeta=\zeta_1+\zeta_2=0.1$,  and the difference in the volume fractions
of the two species $c=\zeta_1-\zeta_2=0.02$. The maximum of $\tilde G_{\alpha\alpha}(k)$ or a minimum of $\tilde G_{12}(k)$ 
is assumed for $k=k_0\approx 0.609$ which corresponds to the minimum of $\tilde V(k)$.
The period of damped oscillations in real space is $\sim 2\piup/k_0\approx 10.3$. In figure~\ref{fig_G(k0)} we show $\tilde G_{\alpha\beta}^\text{co}(k_0)$. Due to the symmetry, $\tilde G_{11}^\text{co}=\tilde G_{22}^\text{co}$ for $c=0$. When $c\ne 0$, the structure factor of the majority component is larger than the structure factor of the minority component, and the difference increases with an increasing asymmetry (see figure~\ref{fig_G(k0)}).
Somewhat surprisingly, the maximum of $\tilde G_{11}^\text{co}(k_0)$ as a function of $c$ is assumed for slightly different volume fractions of the two components, i.e., for $c=0.01$ which corresponds to $\zeta_1=0.055$ and $\zeta_2=0.045$. For larger asymmetries, all the correlations decrease with an increasing $c$. 

Let us focus on the effects of fluctuations. Because of the assumed symmetry of interactions,  we have $\det \tilde{\bf V}(k)=0$,
and from equations~(\ref{Ccc})--(\ref{W'})  we can see that the minimum of $\det \tilde{\bf C}(k)$ is taken for the same value of $k$ as the minimum of $\tilde V(k)$. Thus, $k_0$ is not shifted in this model in the presence of fluctuations. 
Moreover, equation~(\ref{W''}) simplifies to $W''(k_0)=\tilde V''(k_0)P$,
where
\begin{equation}
\label{P}
P=\tilde C_{11}(k_0)+\tilde C_{22}(k_0)+2\tilde C_{12}(k_0).
\end{equation}
The correlation functions in the presence of fluctuations in the Brazovskii-type approximation after some algebra take the forms
\begin{align}
\label{G11}
\tilde G_{11}&=P^{-1}+\frac{[\tilde C_{22}(k_0)+\tilde C_{12}(k_0)]^2}{D(k)}\,, \quad
\tilde G_{22}=P^{-1}+\frac{[\tilde C_{11}(k_0)+\tilde C_{12}(k_0)]^2}{D(k)}\,, \\
\label{G12}
\tilde G_{12}&=P^{-1}-\frac{\tilde C_{11}(k_0)\tilde C_{22}(k_0)+\tilde C_{12}(k_0)[\tilde C_{11}(k_0)+\tilde C_{22}(k_0)+\tilde C_{12}(k_0)]}{D(k)}\,,
\end{align}
where $D(k)=P\lbrace P[ \beta \tilde V(k)-\beta \tilde V(k_0)] +D_0\rbrace $
with $D_0$ defined in (\ref{D0}). In order to calculate the correlation functions, we numerically solved the set of equations (\ref{C(k0)}) for
$\tilde C_{11}(k_0)$, $\tilde C_{22}(k_0)$, and $\tilde C_{12}(k_0)$.

The correlation functions in this approximation, equations~(\ref{G11})--(\ref{G12}), are compared with the 
MF result in figure~\ref{fig_G(k)} for $T^*=0.28$,  $\zeta=0.1$,  and $c=0.02$. As expected, fluctuations lead to a much smaller amplitude and range  of the correlation functions,
especially near the $\lambda$-surface, where  $\tilde C_{\alpha\beta}^\text{co}(k_0)$ diverge in MF. This is manifested by a smaller value and a larger width of  the peak of $\tilde C_{\alpha\beta}(k)$ compared to the MF result. The property that the correlations between particles of the majority component are stronger is still present.
In figure~\ref{fig_G(k0)} we show  $\tilde G_{11}(k_0)$ for $T^*=0.28$, $\zeta=0.1$ and for a range of $c$,
with the fluctuation contribution included. The nonmonotonous dependence on $c$ is enhanced in the presence of fluctuations, and the maximum occurs for $c=0.02$. The magnitude of the other correlation functions decreases monotonously with an increasing $c$. 

\begin{figure}[!t]
	\includegraphics[scale =0.35]{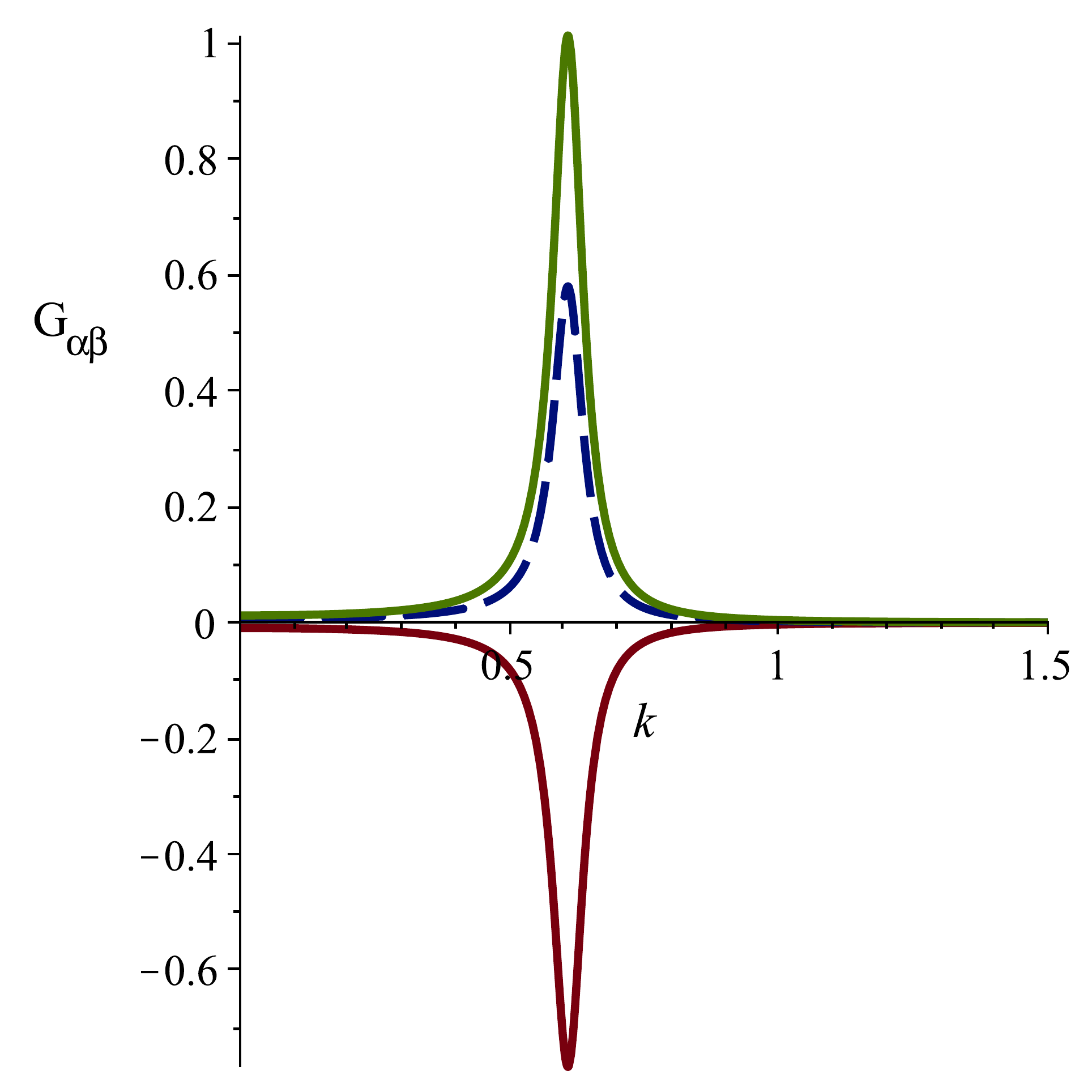}
	\includegraphics[scale =0.35]{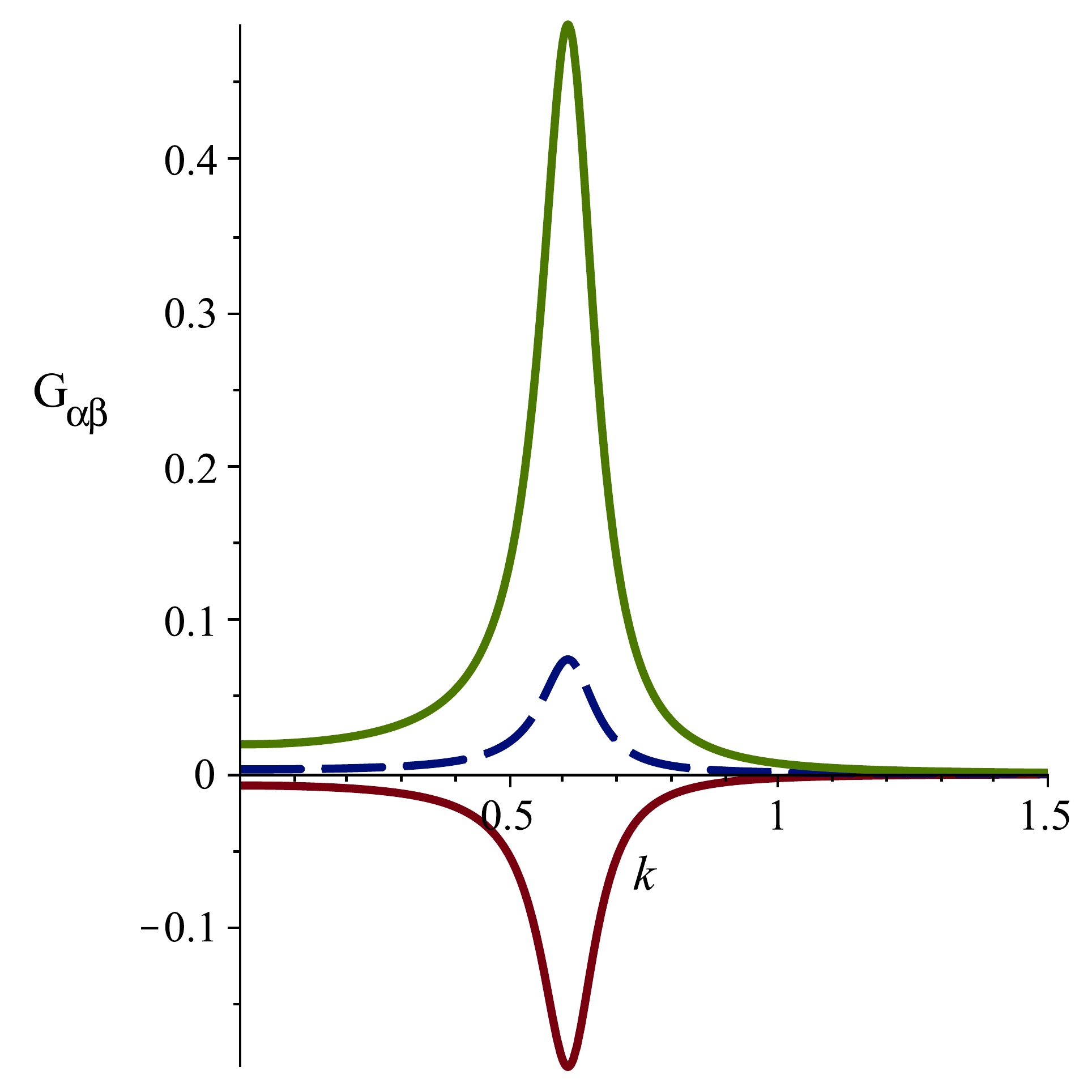}
	\caption{(Colour online) Correlation functions in Fourier representation  for $T^*=0.12$ and $\zeta=0.1$ with the effect of fluctuations taken into account. Upper solid lines: $\tilde G_{11}(k)$,  dashed lines: $\tilde G_{22}(k)$ and lower solid lines: $\tilde G_{12}(k)$. Left-hand panel:  $c=0.02$. Right-hand panel: $c=0.06$.
	}
	\label{figG(k)T1}
\end{figure}

The peak of the correlation functions becomes higher and narrower for an increasing density and/or a decreasing temperature, signaling correlations stronger and of larger range. The effect of temperature can be seen by comparing figure~\ref{fig_G(k)} 
for $T^*=0.28$ with  figure~\ref{figG(k)T1} for $T^*=0.12$, both for $\zeta=0.1$. In figure~\ref{figG(k)T1}, the correlation functions  are shown  for $c=0.02$ and $c=0.06$.  We can see a strong 
effect of the difference in concentrations of the species $1$ and $2$ on the correlation functions. All correlations decrease, but especially the minority component becomes much less correlated when its volume fraction decreases. 

\begin{figure}[!t]
	\centering
	\includegraphics[clip,width=0.40\textwidth,angle=0]{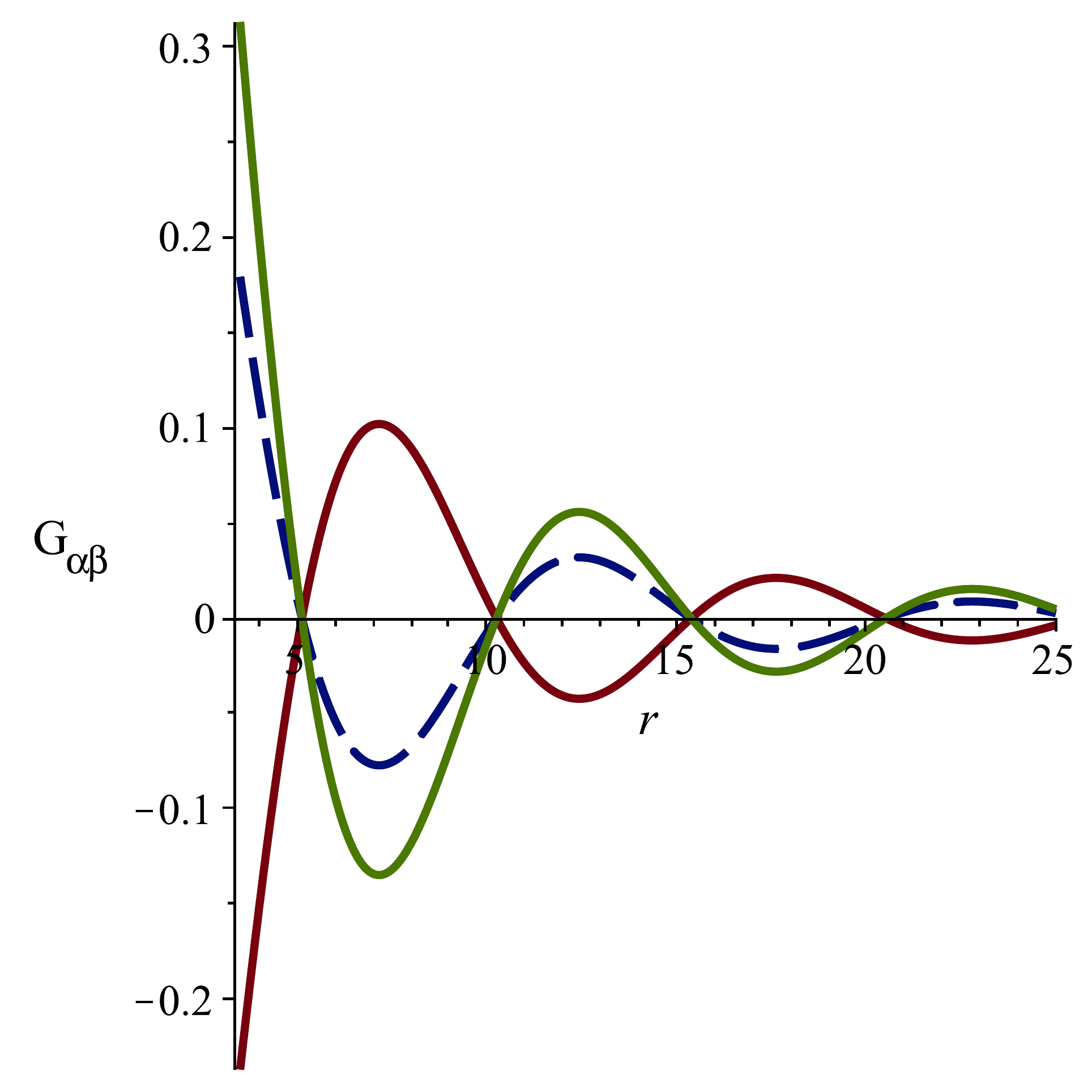}
	\caption{(Colour online) Correlation functions in real space for $T^*=0.12$, $\zeta=0.1$ and $c=0.02$ with the effect of fluctuations taken into account.
		$ G_{11}(r)$ (green upper solid line on the left), $ G_{22}(r)$ (dashed line) and $G_{12}(r)$ (red lower solid line on the left).}
	\label{fig_G(r)}
\end{figure}

It is also interesting to consider the correlation function for the total volume fraction, 
$G({\bf r}_1-{\bf r}_2)=\langle \Delta\zeta({\bf r}_1)\Delta\zeta({\bf r}_2)\rangle$. Equations~(\ref{G11})--(\ref{G12}) give
\[
\label{Gtot}
\tilde G(k)=4P^{-1}+\frac{\left[ \tilde C_{22}(k_0)-\tilde C_{11}(k_0)\right]^2}{D(k)}.
\]
This equation and figure~\ref{figG(k)T1} show that  the correlations for the total volume fraction increase with an increasing asymmetry which in this case is induced by increasing $c$. In the fully symmetrical case, $\tilde G(k)$ vanishes in this theory.
If the assumptions leading to the fluctuation contribution obtained in the Brazovskii-type approximation are valid, we can make the approximation
\begin{equation}
\tilde V(k)-\tilde V(k_0)\approx \frac{V''(k_0)}{8k_0^2}(k^2-k_0^2)^2\,,
\label{Delta_V}
\end{equation}
where we took into account that $\tilde G_{\alpha\beta}(k)$ should be even functions of $k$.
In this approximation, the correlation functions in real-space representation can be calculated analytically by inverse Fourier transforming equations~(\ref{G11})--(\ref{G12}). The results
have the form (\ref{Gr}) with $\theta=0$, and the inverse lengths $\alpha_0,\alpha_1$ are given by
\[
2\alpha_{0}^2=\left\lbrace  8k_0^2\left[\frac{k_0^2}{8}+\frac{T^*D_0}{P \tilde{V}''(k_0)}\right] \right\rbrace  ^{1/2}-k_0^2\,, \qquad
2\alpha_1^2=\left\lbrace  8k_0^2\left[ \frac{k_0^2}{8}+\frac{T^*D_0}{P\tilde V''(k_0)}\right] \right\rbrace  ^{1/2}+k_0^2\,,
\]
where $P$ and $D_0$ are defined in (\ref{P}) and  (\ref{D0}), respectively.
In figure~\ref{fig_G(r)} we show the correlation functions in real space for $T^*=0.12$,  $\zeta=0.1$ and $c=0.02$.
The negative extremum of $\tilde C_{12}(k)$ at $k=k_0$ leads to the opposite signs of $C_{12}(r)$ and $C_{\alpha\alpha}(r)$ for the same value of $r$. This means that the regions enriched in one component are depleted in the other component. 

\subsection{Simulations}
A binary mixture of particles interacting with the SALR potential 
(\ref{int_pot_r}) with  $K_1=1,K_2=0.2,\kappa_1=~1,
\newline \kappa_2=0.5$
was simulated by applying a Monte Carlo technique in the $NVT$ ensemble.  
The particles had the same diameter ($\sigma_{1}=\sigma_{2}=\sigma= 1.0$), and were placed in a cubic box of
edge length $50\sigma$, with periodic boundary conditions  applied to the system in the three directions.
Cut-off radius was  $15\sigma$. Each system had run $10^{6}$ Monte Carlo steps for equilibration and 
$10^{5}$ for production.
\begin{figure}[!b]
	\centering
	\includegraphics[clip,width=0.42\textwidth,angle=0]{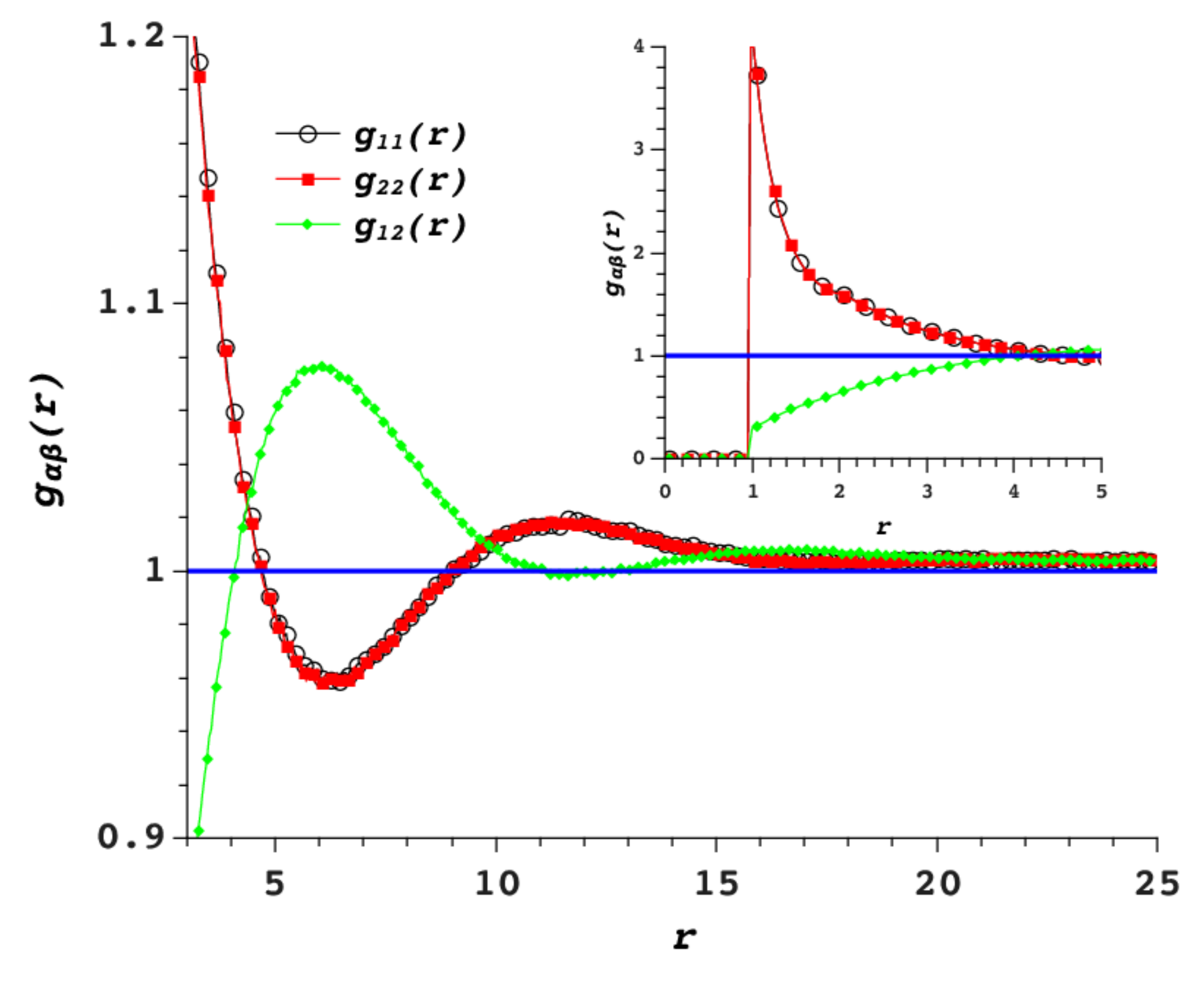}
	\includegraphics[clip,width=0.42\textwidth,angle=0]{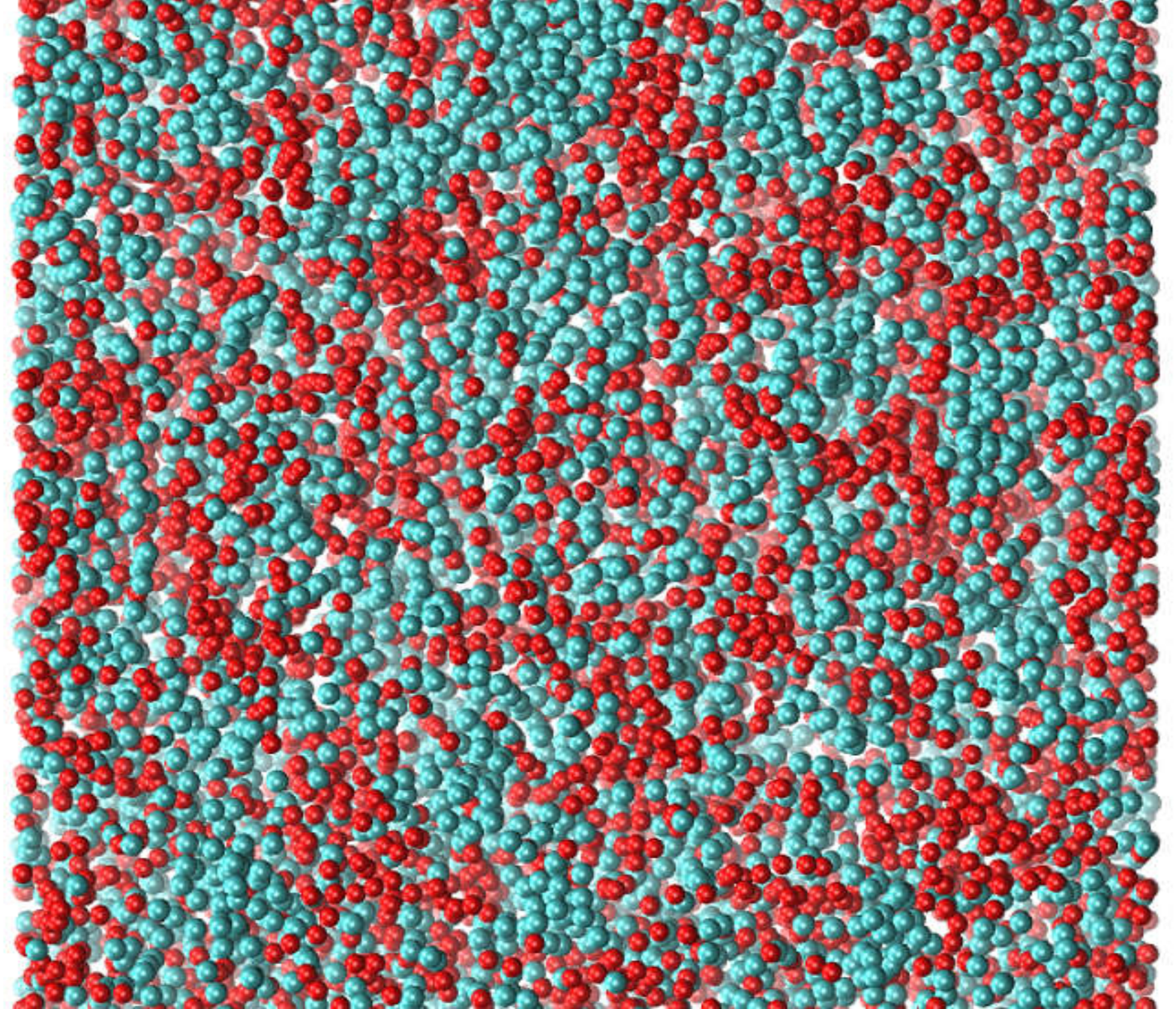}
	\caption{(Colour online) Pair distribution functions and a representative configuration for  $T^*=0.28$, and volume fractions 
		$\zeta_1=\zeta_2=0.026$. }
	\label{sim1}
\end{figure}

\begin{figure}[!t]	
	\centering
	\includegraphics[clip,width=0.44\textwidth,angle=0]{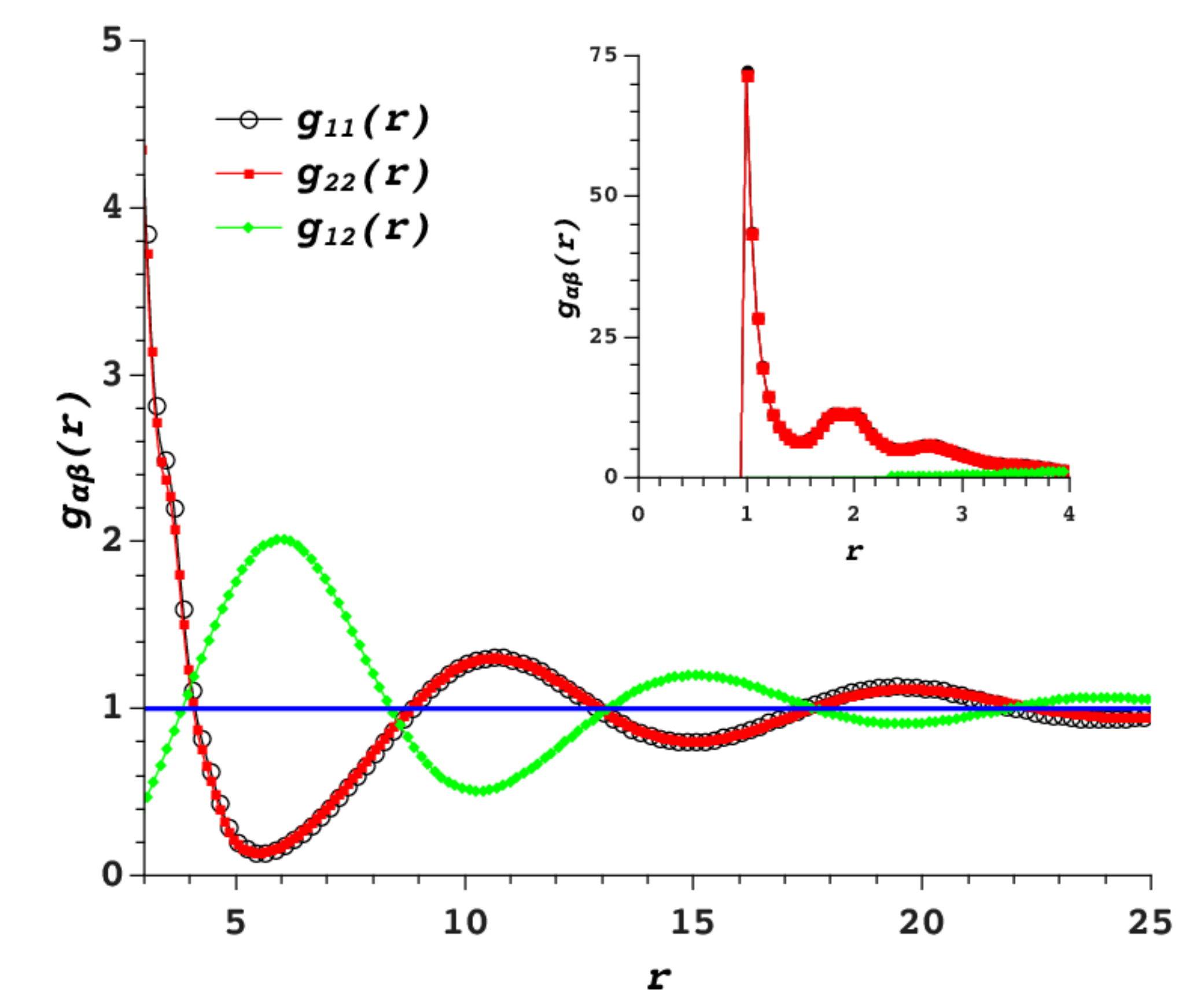}
	\includegraphics[clip,width=0.44\textwidth,angle=0]{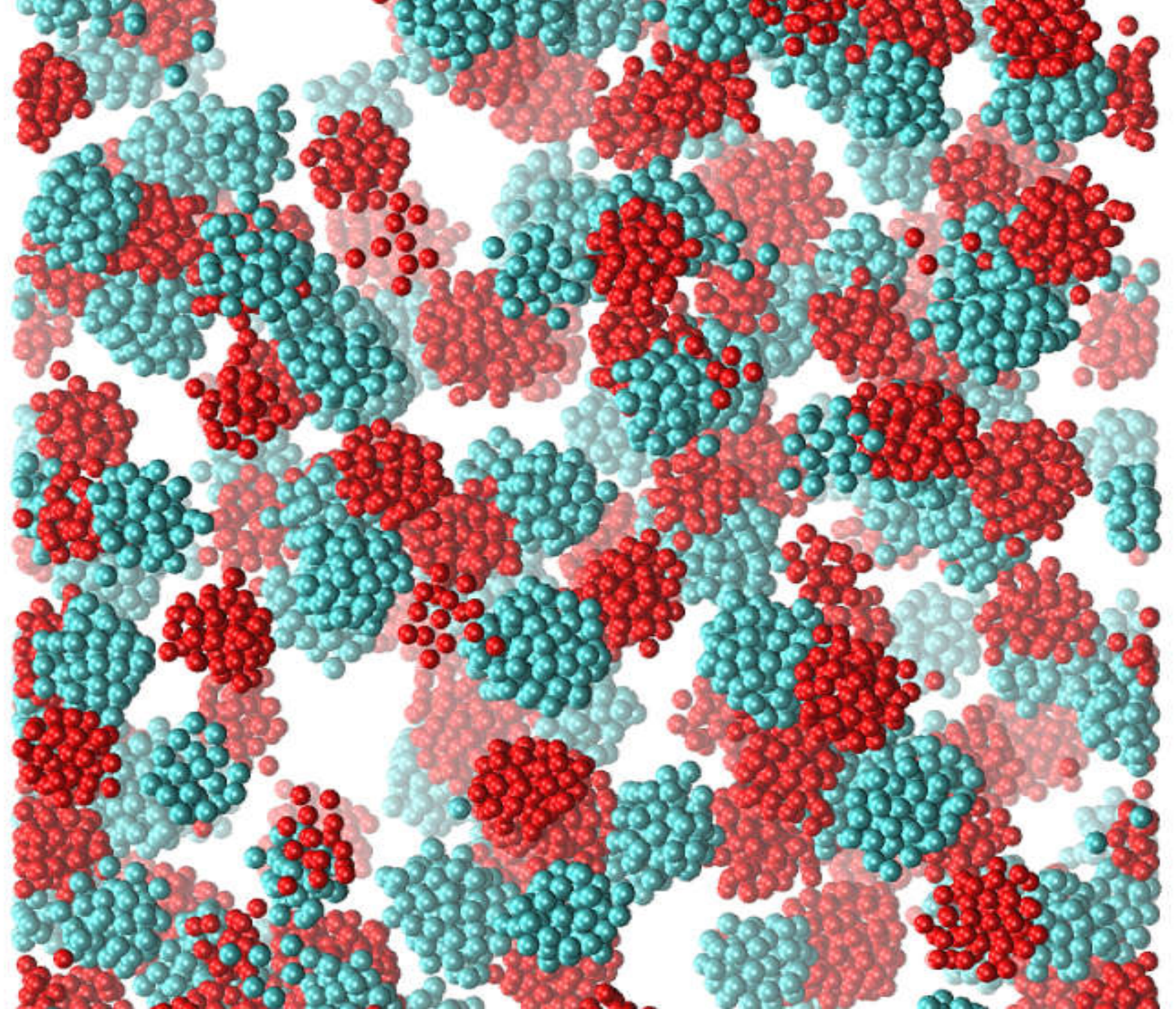}
	\caption{(Colour online) Pair distribution functions and a representative configuration  for $T^*=0.12$, and volume fractions 
		$\zeta_1=\zeta_2=0.026$.  }
	\label{sim2}
\end{figure}

We present the pair distribution functions,
$g_{\alpha\beta}=\frac{G_{\alpha\beta}}{\zeta_{\alpha}\zeta_{\beta}}+1$,
and  representative configurations for $\zeta=0.052$ and $c=0$
in figures~\ref{sim1} and \ref{sim2} for   $T^*=0.28$,  and  $T^*=0.12$, respectively.
In both cases, an oscillatory decay with the period of damped oscillations $\lambda\approx 10$ can be seen, but the 
amplitude is much larger for $T^*=0.12$. This larger amplitude and range of correlations is reflected in much more 
ordered configuration shown in figure~\ref{sim2}, as compared with the configuration shown in figure~\ref{sim1}. 
At $T^*=0.12$, we can see the chains of well assembled clusters, where clusters of different species are attached to each other. 
At $T^*=0.28$, the inhomogeneities are much less visible.

\begin{figure}[!t]
		\centering
	\includegraphics[clip,width=0.4\textwidth,angle=0]{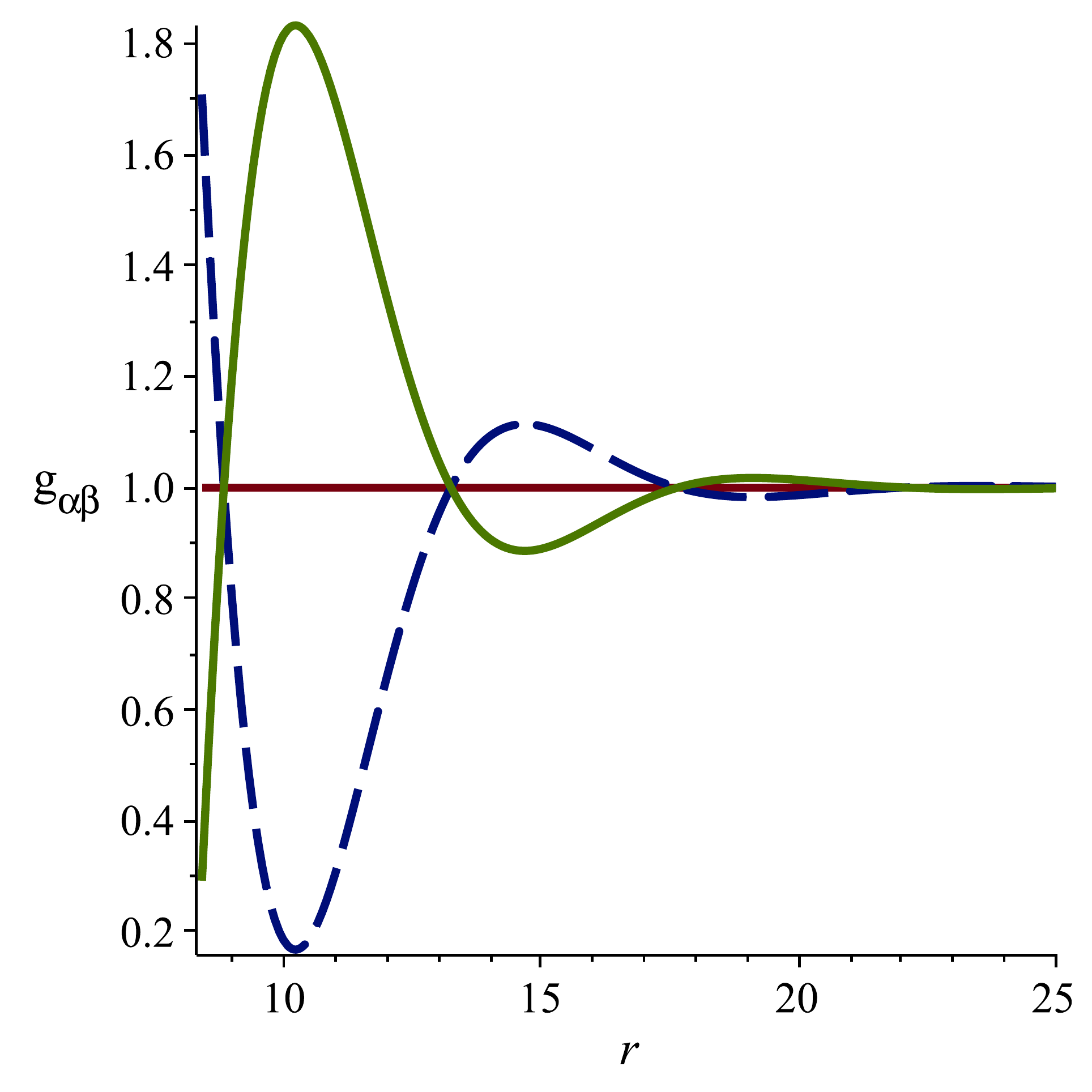}
	\includegraphics[clip,width=0.4\textwidth,angle=0]{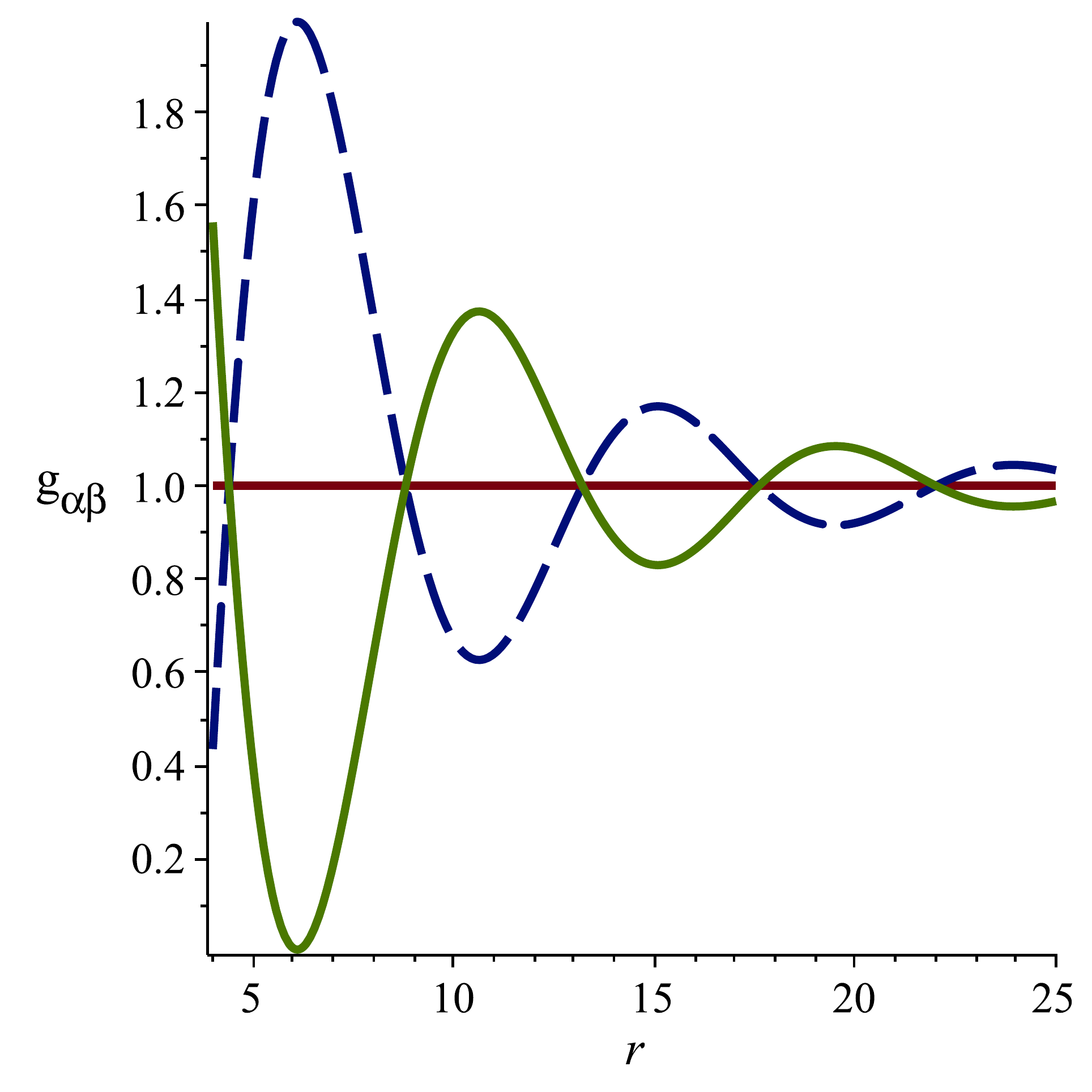}
	\caption{(Colour online) Left-hand panel: pair distribution functions in real space for $T^*=0.12$, $\zeta_1=\zeta_2=0.026$ 
		with the effect of fluctuations taken into account in the mesoscopic theory.
		$ g_{11}(r)=g_{22}(r)$ (solid line), and $g_{12}(r)$ (dashed line). Right-hand panel: Functions $g_{\alpha\beta}=G_{\alpha\beta}+1$, where $G_{\alpha\beta}$ is
		given in equation~(\ref{Gr})
		with $\alpha_1=0.714$, $\alpha_0=0.1$, $\theta=0$ and $A=\pm 8.5$.  
	}
	\label{fig_g(r)}
\end{figure}

We have computed $\tilde G_{\alpha\beta}$ for the thermodynamic states corresponding to figures~\ref{sim1} and \ref{sim2}. 
In the first case, we obtained a quite flat maximum of $\tilde G_{\alpha\alpha}(k)$. 
As far as a pronounced maximum is assumed in our theory and the theory should be self-consistent, 
for this thermodynamic state our theory is not accurate enough. In the second case, the maximum of 
$\tilde G_{\alpha\alpha}(k)$ is  better developed, but the peak is not as high and narrow as for $\zeta=0.1$ at the same $T^*$.
The corresponding pair distribution functions  are shown in figure~\ref{fig_g(r)} for $r>8$, 
because in the mesoscopic theory the results for the correlation function cannot be valid for distances smaller than one
period of the damped oscillations.

By comparing figure~\ref{fig_g(r)} with the simulation results (figure~\ref{sim2}) we can see that in the theory 
the amplitude is too large and the decay length is too small. In figure~\ref{fig_g(r)} (right-hand panel)
we plot $g_{\alpha\beta}=G_{\alpha\beta}+1$, where $G_{\alpha\beta}$ is  given in (\ref{Gr}) with $\alpha_1$
determined in our theory, but with $\alpha_0=0.1$ about 5 times smaller than predicted by the theory.
The amplitude $A=8.5$ in equation~(\ref{Gr}) for
$g_{11}(r)$ and $A=-8.5$ for $g_{12}(r)$ is also much smaller. We did not try to find the best fit
to the simulation results, but it is
clear that for the chosen parameters the agreement between simulations and equation~(\ref{Gr}) is very good. From this agreement, it immediately follows that in simulations the structure factor for the like (different) particles should assume a maximum (minimum) for $k\approx k_0$, and near the extremum it behaves as $|S_{\alpha\beta}|\propto [\text{const} +(k^2-k_0^2)^2]^{-1}$.
The theory developed in this work is based on the Brazovskii approximation. In ~\cite{ciach:16:1},
additional correction to the MF inverse correlation function is taken into account. This 
correction term in one-component systems is proportional to $-A_3(\zeta)^2$, and leads to a smaller value of 
the inverse correlation function than in the Brazovskii approximation. Hence, larger correlations can be expected.
A satisfactory agreement with the exact results 
was obtained in~\cite{ciach:16:1} for an equation of state in a one-dimensional model when the
correction  proportional to $-A_3(\zeta)^2$
was taken into account. Since $A_3(\zeta)=0$ for the critical volume fraction $\zeta\approx 0.129$, the Brazovskii
approximation works well for $\zeta\approx 0.1$. $A_3(\zeta)$ increases to large values for $\zeta$ decreasing 
from $\zeta\approx 0.129$,
and the accuracy of the Brazovskii approximation should decrease for a decreasing $\zeta$. Our results show that for
$\zeta\approx 0.05$, the Brazovskii-type approximation is not sufficiently accurate. 
The period of damped oscillations, however, remains close to $2\piup/k_0$, and the formula (\ref{Gr}) is still valid. 
\begin{figure}[!t]
	\centering
	\includegraphics[clip,width=0.44\textwidth,angle=0]{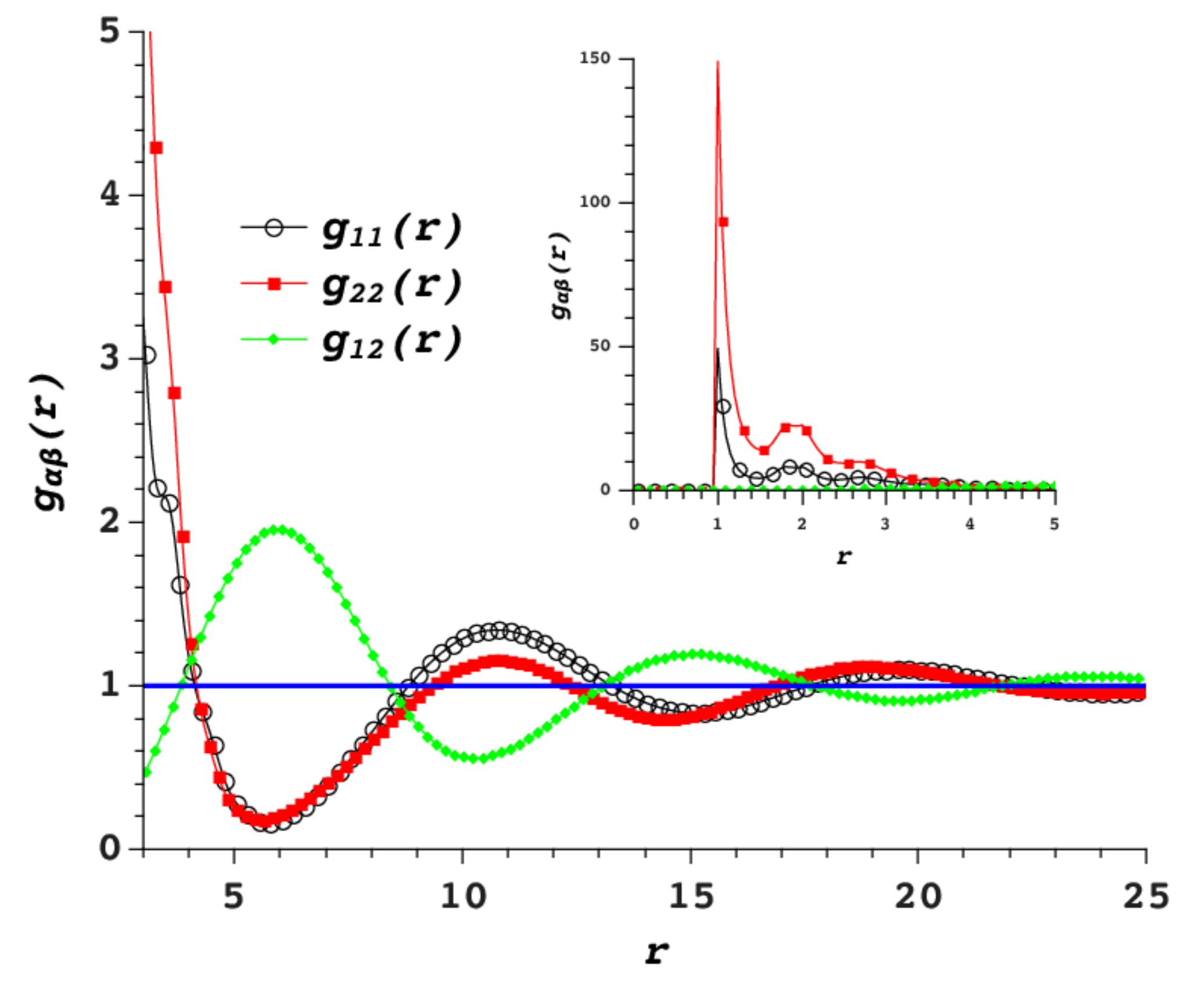}
	\includegraphics[clip,width=0.44\textwidth,angle=0]{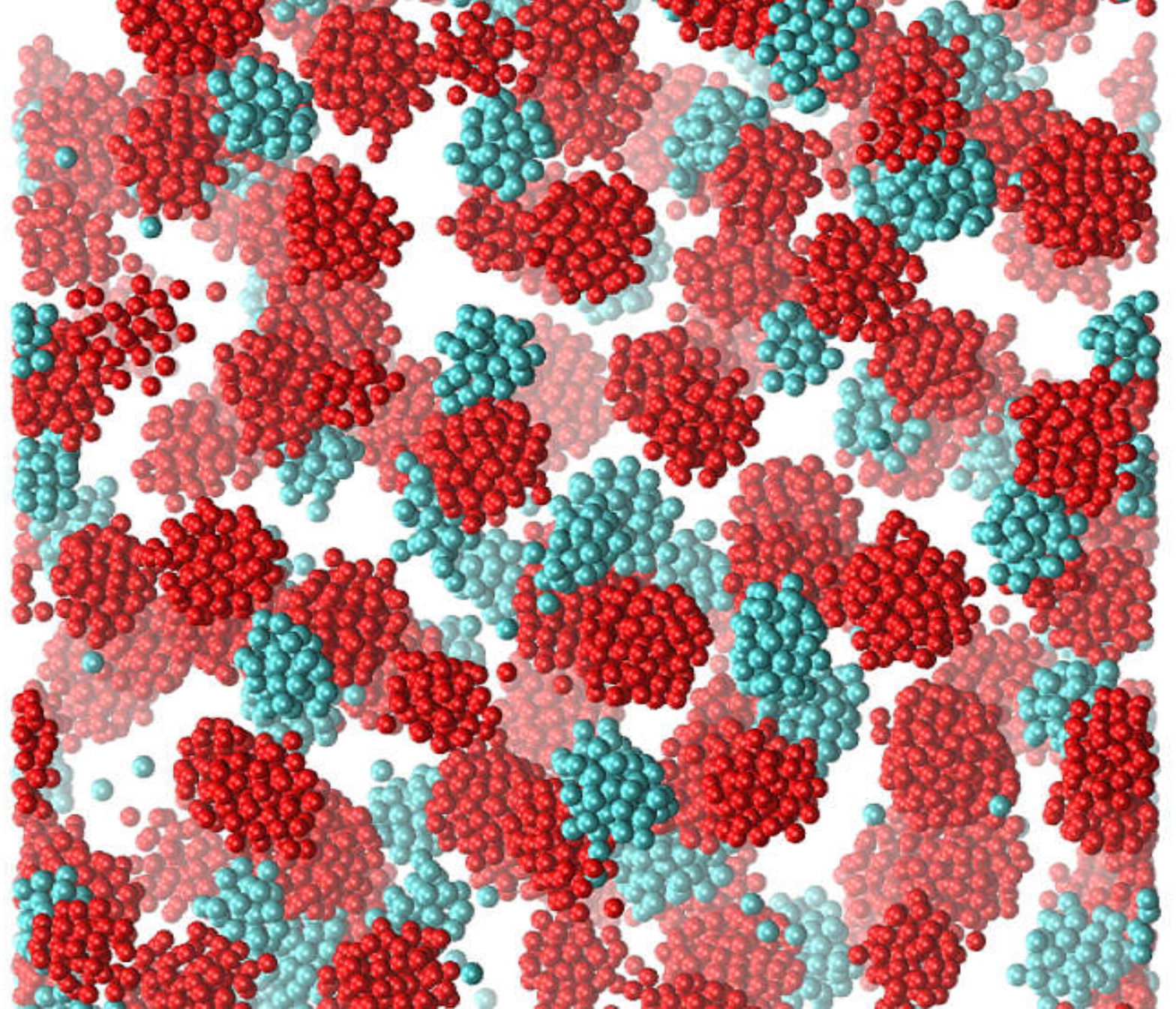}
	\caption{(Colour online) Pair distribution functions and a representative configuration  for $T^*=0.12$, and volume fractions 
		$\zeta_1=0.042$, $\zeta_2=0.01$. }
	\label{sim3}
\end{figure}

The effect of asymmetry of the volume fractions of the two species is shown in figure~\ref{sim3}  for $T^*=0.12$,
$\zeta=0.052$ and $c=0.032$. Correlations between the majority species are stronger,
in agreement with theoretical predictions. Clusters are of a similar size as in the symmetrical case.
However, there are not enough particles of the minority component to fill the space between two clusters of 
the majority component, 
and this space remains empty. Recall that due to the long-range repulsion between the like particles,
clusters of the same kind repel each other. For this reason, the correlation function for total volume 
fraction oscillates, in contrast to the symmetrical case. This result also agrees with theoretical predictions. 
\section{The case of interaction  potentials  $V_{11}=V_{22}=0$ and $V_{12}\ne 0$}\label{sec4}

Now, we consider a particular case of the model binary mixture in which  there are only hard-core repulsions between the like-particles while the particles of different types interact through the attractive  potential $V_{12}(r)$ beyond the hard core. The mixture can be perceived as a crude model of charged colloid particles in the solvent containing depletion agents.  When charges are switched off, colloid particles  attract each other due to  the presence of depletants. If the particles are charged, screened electrostatic attraction between different particles appears, while  the like-particles repel each other. For a suitably chosen size of the depletant and the screening length and charge, the attraction and repulsion between the like-particles can cancel each other approximately, and the attraction between the particles of different types can be doubled.

\subsection{Theory}
\begin{figure}[!t]
	\includegraphics[width=0.4\textwidth]{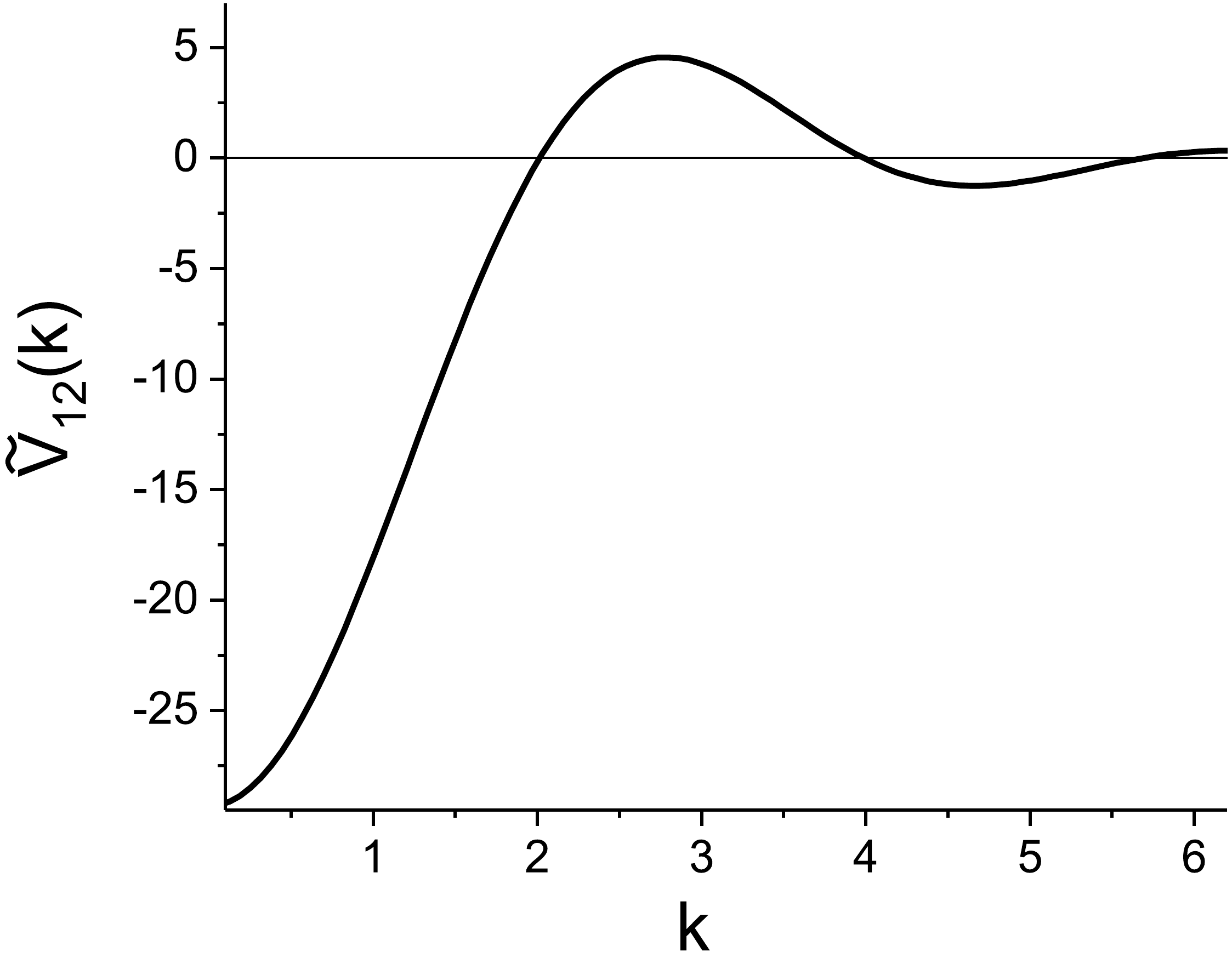}
	\hfill	
	\includegraphics[width=0.4\textwidth]{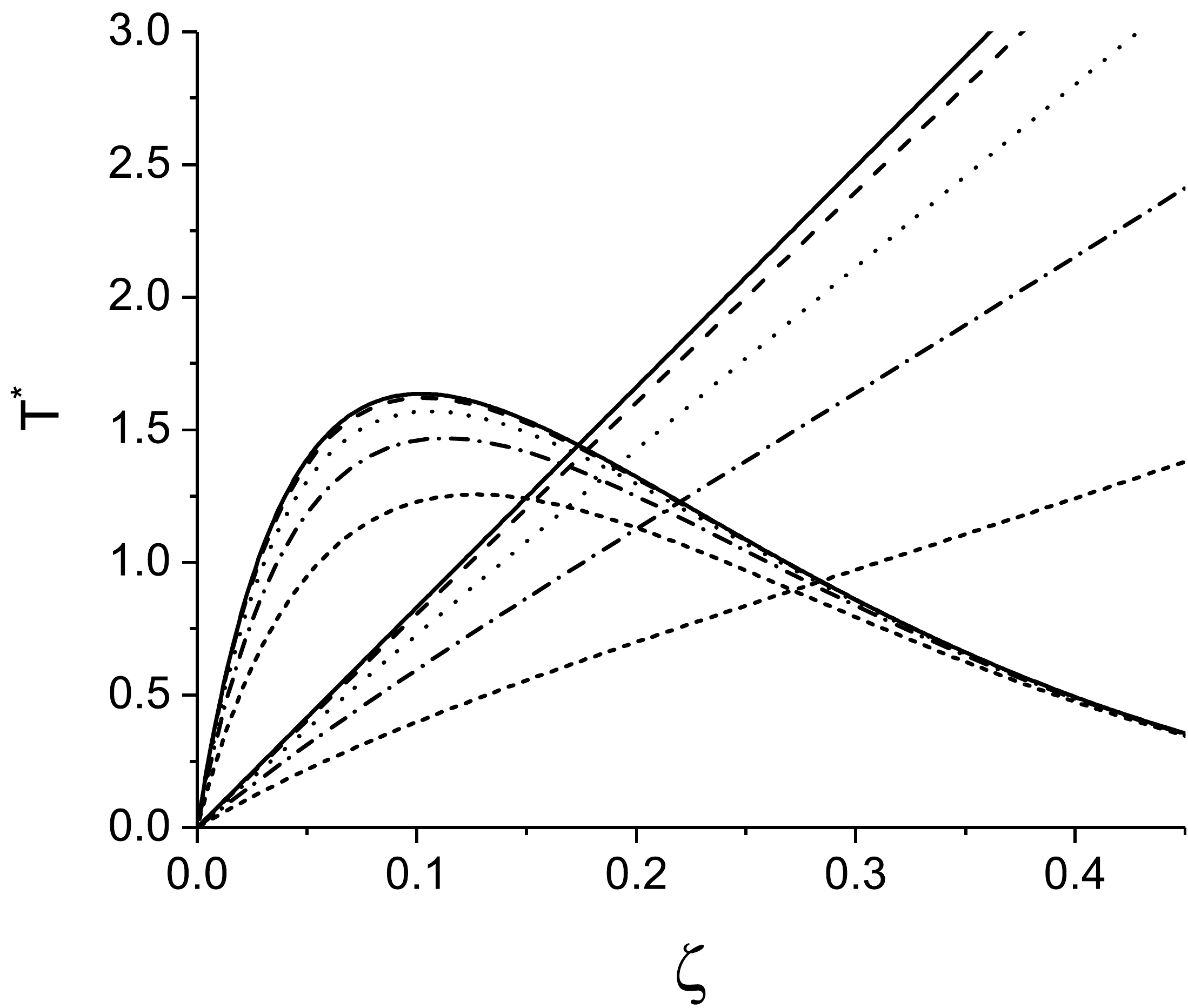}	
	\\
	\parbox[t]{0.47\textwidth}
	{\caption{Interaction potential between particles of different kinds (equation (\ref{V_sq-well}) for $a=2$) in Fourier representation. The wave-number $k$ is in $\sigma^{-1}$ units.}
		\label{V12-k}}
	\hfill
\parbox[t]{0.47\textwidth}
	{\caption{
			Gas-liquid spinodals and $\lambda$-lines for the model (\ref{V11-V22})--(\ref{V_sq-well}) at $a=2$ for different  $c^*=(\zeta_{1}-\zeta_{2})/\zeta$:  $c^*=0$ (solid lines), $0.2$ (dashed lines), $0.4$ (dotted lines), $0.6$ (dash-dotted lines), and $0.8$  (short-dashed lines). 
		}\label{spinodals}}
\end{figure}

As the first step, we assume equal sizes of particles for all species.  
For $V_{12}(r)$, we choose the  square well potential. 
Thus, the model is characterized by the following interaction potentials beyond the hard core
\begin{align}
V_{11}(r)&=V_{22}(r)=0, 
\label{V11-V22}
\end{align} 
\begin{align}
V_{12}(r)&=-\varepsilon\theta(r-1)\theta(a-r),
\label{V_sq-well}
\end{align}
where $a>1$ is the range of  the potential, $\varepsilon$ is the interaction strength at contact of the two unlike 
particles and $r$ is  in  $\sigma$ units. We are interested in periodic structure on the length scale of $\sigma$, not in any particular system. Since in \cite{ciach:11:2} the MF phase diagram was obtained for $a=2$, we assume $a=2$ here too.
The Fourier transform of  the potential (\ref{V_sq-well}) for $a=2$ 
is shown in figure~\ref{V12-k}.
The MF analysis showed \cite{ciach:11:2} that the model undergoes two types of instability: one connected with $k=0$ and the other one with $k_{0}\approx 2.78$. The former is related to  the gas-liquid phase separation which occurs at lower volume fractions while the latter
is related to the appearance of  local inhomogeneity at the length scale $2\piup/k_{0}$.  In this model, inhomogeneous structures can occur when the Fourier transform of the interaction potential between unlike particles has positive maximum for $k>0$.  

First, we focus on the MF approximation. In this case, the correlation functions in Fourier representation obtained for the model are of the form:
\begin{eqnarray}
\tilde G_{11}^\text{co}(k)=\frac{A_{22}}{D^\text{co}(k)}\,,  \quad
\tilde G_{22}^\text{co}(k)=\frac{A_{11}}{D^\text{co}(k)}\,,
\quad
\tilde G_{12}^\text{co}(k)=-\frac{A_{12}+\beta\tilde{V}_{12}(k)}{D^\text{co}(k)}\,,
\label{G12-mf-sw}
\end{eqnarray}
where
$D^\text{co}(k)=A_{11}A_{22}-\left[ A_{12}+\beta\tilde{V}_{12}(k)\right]^{2}$
and $\tilde{V}_{12}(k)$  is the Fourier transform of the interaction potential~(\ref{V_sq-well}). From  the equation $D^\text{co}(k)=0$ one can  get the expressions for  the gas-liquid spinodal  and for the $\lambda$-surface, respectively
\begin{eqnarray}
T_\text{sp}^*=\frac{\tilde{V}_{12}(0)}{\sqrt{A_{11}A_{22}}+A_{12}}\,, \qquad
T_{\lambda}^*=\frac{\tilde{V}_{12}(k_0)}{\sqrt{A_{11}A_{22}}-A_{12}}\,,
\label{lambda-line} 
\end{eqnarray}
where the dimensionless temperature is defined as $T^*=k_\text{B}T/\varepsilon$. In figure~\ref{spinodals}, we present the $T^*$--$\zeta$-plots of the MF boundaries of stability  determined by   equations (\ref{lambda-line})  for different values of $c^*=(\zeta_{1}-\zeta_{2})/\zeta$.
\begin{figure}[!t]
	\centering	\includegraphics[clip,width=0.42\textwidth,angle=0]{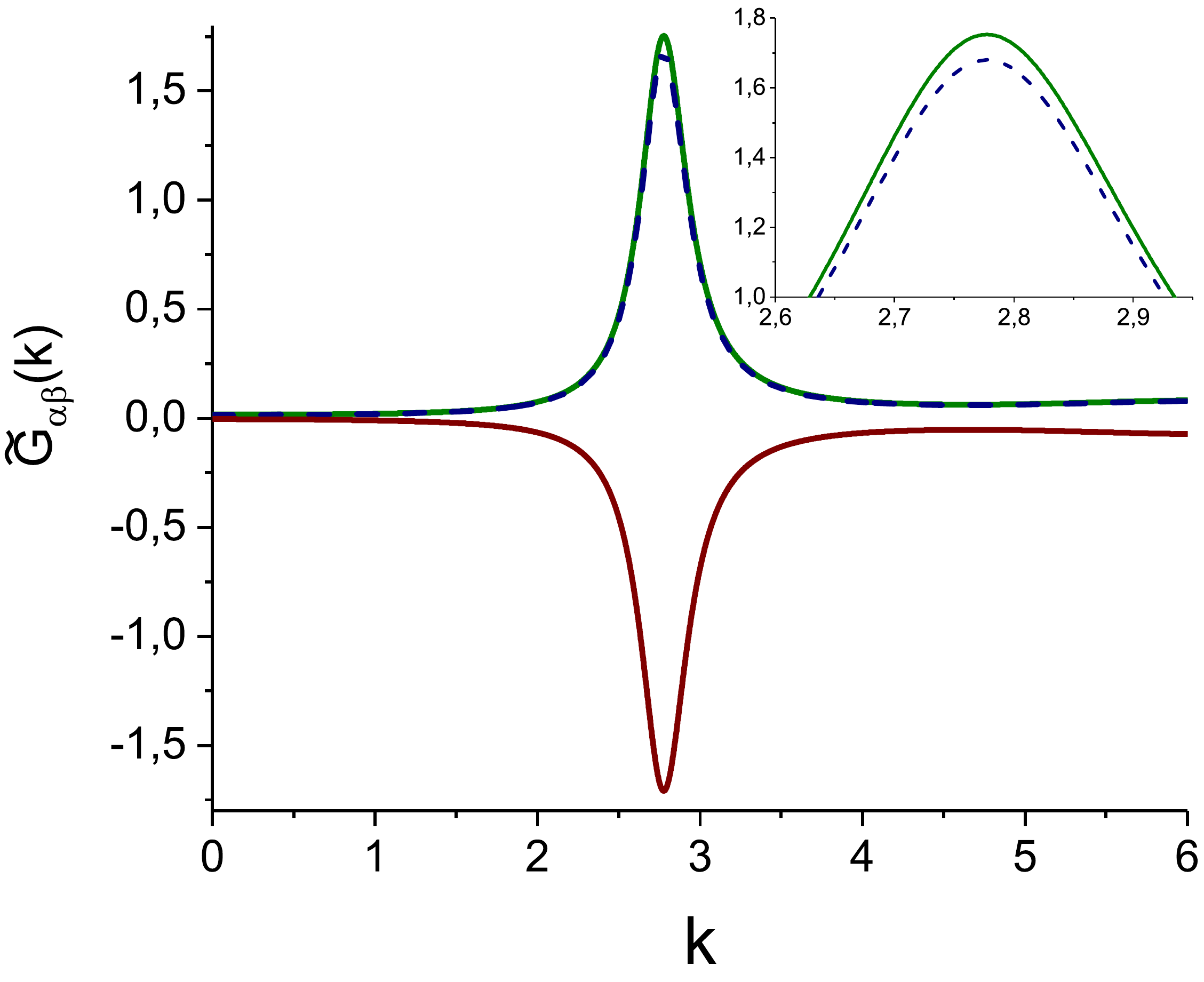} \qquad
	\includegraphics[clip,width=0.42\textwidth,angle=0]{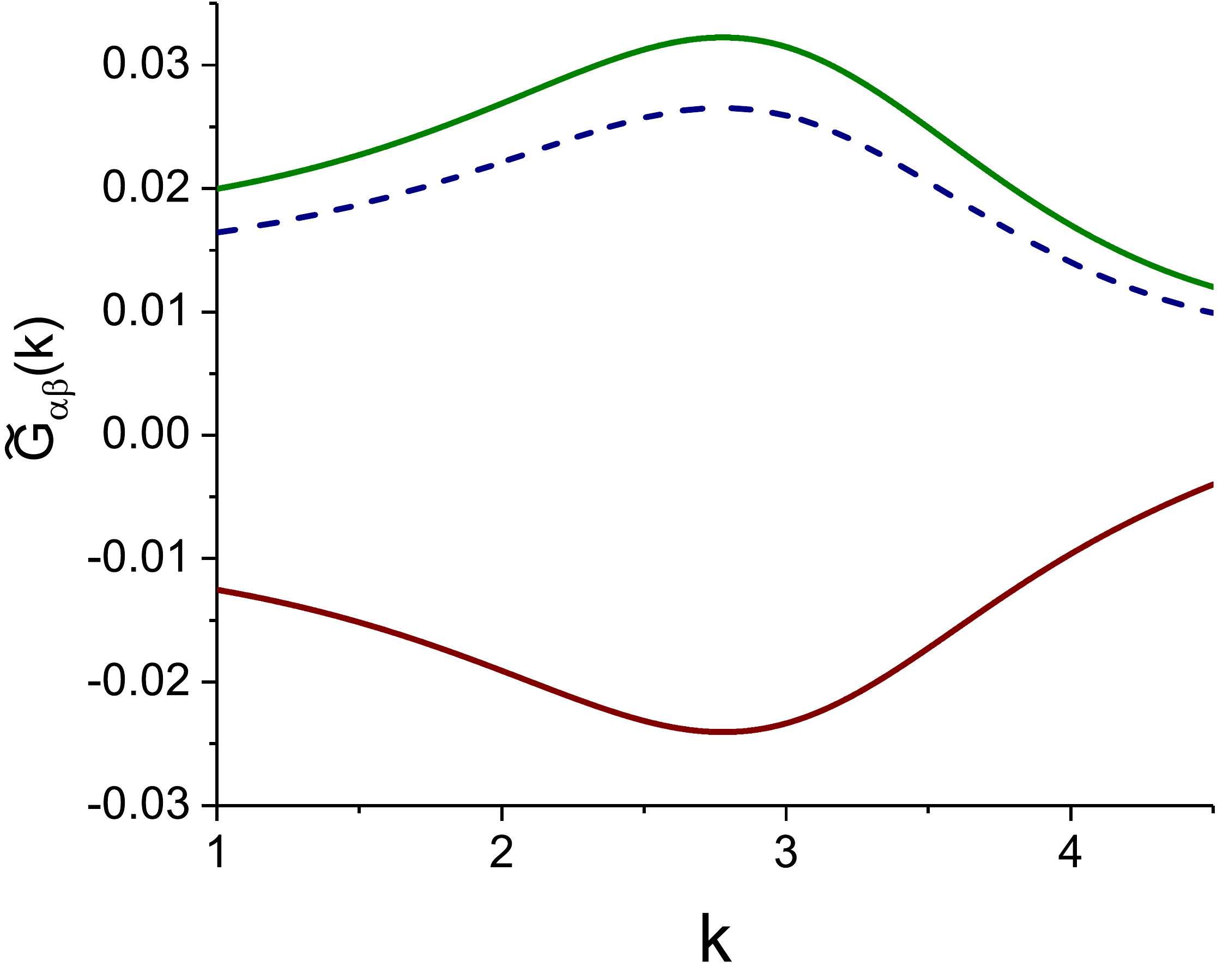}
	\caption{\label{GIJ-mf(k)}
	(Colour online)	Correlation functions  in Fourier representation  for $T^*=2.5$, $\zeta=0.3$ and $c=0.06$ in MF approximation.
		The inset magnifies the region between $k=2.6$ and $k=3$ (left-hand panel) and with the effect of fluctuations 
		taken into account (right-hand panel). Upper solid lines: $\tilde G_{11}(k)$,   dashed lines: $\tilde G_{22}(k)$, 
		and  lower solid lines: $\tilde G_{12}(k)$.
	}
\end{figure}

Using equations 
(\ref{G12-mf-sw}), we calculate the MF correlation functions $\tilde{G}^\text{co}_{\alpha\beta}(k)$ above the $\lambda$-surface. Due to  the proximity of the gas-liquid phase separation we chose the volume fraction   equal to $0.3$ which is  higher than in the case of the `two mermaids and a peacock'  model. As before, we assume that the majority component is the species 1, and consider only $c\geqslant 0$. The correlation functions
$\tilde{G}^\text{co}_{\alpha\beta}(k)$ are shown in figure~\ref{GIJ-mf(k)} (left-hand panel) for $T^{*}=2.5$ and $c=0.06$. The main maximum (minimum) of $\tilde{G}^\text{co}_{\alpha\beta}(k)$ corresponds to the maximum of the interaction potential $\tilde{V}_{12}$ (see figure~\ref{V12-k}). We observe  a very small difference in the  peak heights of $\tilde{G}^\text{co}_{11}(k)$ and $\tilde{G}^\text{co}_{22}(k)$ in this case.
\begin{figure}[!t]
	\centering
	\includegraphics[clip,width=0.42\textwidth,angle=0]{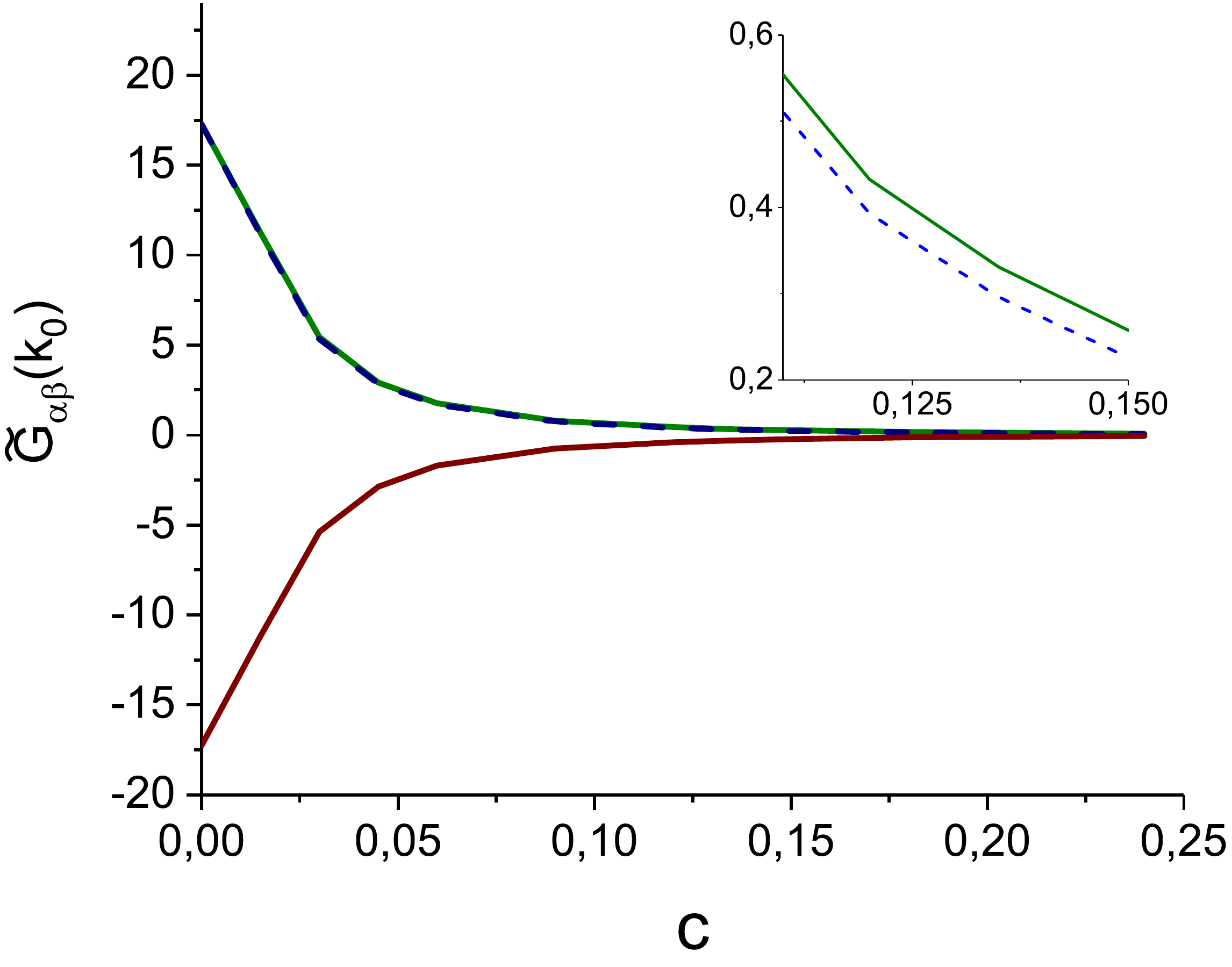} \qquad
	\includegraphics[clip,width=0.42\textwidth,angle=0]{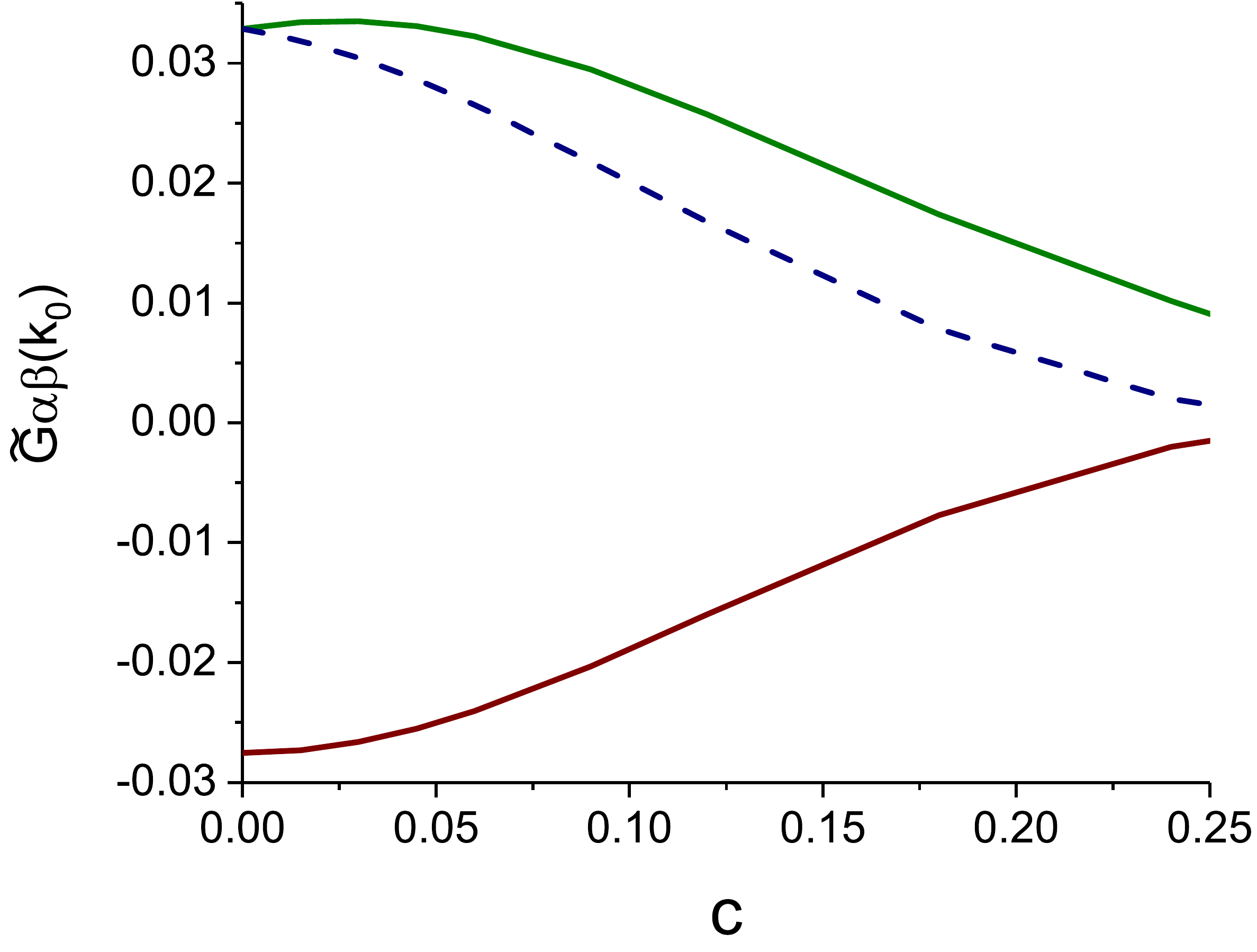}
	\caption{\label{GIJ_c(k0)}
		(Colour online) Left-hand panel: $\tilde G_{11}(k_0)$ (upper solid line), $\tilde G_{22}(k_0)$ (dashed line) and $\tilde G_{12}(k_0)$ (lower solid line)    in the MF approximation. The inset magnifies the region between $c=0.11$ and $c=0.15$. 
		Right-hand panel:  $\tilde G_{11}(k_0)$ and $\tilde G_{22}(k_0)$   with the effect of fluctuations taken into account.    $T^*=2.5$, $\zeta=0.3$ and
		$c=\zeta_{1}-\zeta_{2}$.
	}
\end{figure}
In figure~\ref{GIJ_c(k0)} (left-hand panel), we show the dependence of  $\tilde G^\text{co}_{\alpha,\beta}(k_{0})$ on $c$. We see a monotonous  dependence of the three correlation functions  on $c$. At the same time, the difference in the peak heights of   $\tilde G^\text{co}_{11}(k_0)$ and $\tilde G^\text{co}_{22}(k_0)$ increases a bit with an increase of $c$. 

Now we consider the effect of fluctuations on the correlation functions. Like the `two mermaids and a peacock' model considered in the previous section,  $k_{0}$ in our model is not shifted when the fluctuations are included. For  this model, 
$W''(k_{0})=-2\beta\tilde{V}_{12}''(k_{0})\tilde{C}_{12}(k_{0})$.
As a result, equations (\ref{C(k)}) reduce to the form:
\begin{align}
\tilde{C}_{\alpha\alpha}(k)&=\tilde{C}_{\alpha\alpha}(k_{0}), \quad \alpha=1,2, \nonumber\\
\tilde{C}_{12}(k)&=\tilde{C}_{12}(k_{0})+ \beta\Delta\tilde{V}_{12}(k),
\label{C12_fl}
\end{align}
where  $\Delta\tilde{V}_{12}(k)=\tilde{V}_{12}(k)-\tilde{V}_{12}(k_{0})$. The  correlation functions $\tilde{C}_{\alpha\beta}(k_{0})$ are obtained by solving equations~(\ref{C(k0)}). Finally,
correlation functions $\tilde{G}_{\alpha\beta}(k)$ can be presented as follows:
\begin{eqnarray}
\tilde{G}_{11}(k)=\frac{\tilde{C}_{22}(k_{0})}{D(k)}\,, \quad
\tilde{G}_{22}(k)=\frac{\tilde{C}_{11}(k_{0})}{D(k)}\,, \quad
\tilde{G}_{12}(k)=-\frac{\tilde{C}_{12}(k_{0})+\beta\Delta\tilde{V}_{12}(k)}{D(k)}\,,
\label{G12-fl-sw}
\end{eqnarray}
where  $D(k)=\tilde{C}_{11}(k_{0})\tilde{C}_{22}(k_{0})-[ \tilde{C}_{12}(k_{0})+\beta\Delta\tilde{V}_{12}(k)]^{2}$.
In the above equations, for $\Delta\tilde{V}_{12}(k)$ we use the approximation (\ref{Delta_V}).
\begin{figure}[!t]
	\centering
	\includegraphics[clip,width=0.45\textwidth,angle=0]{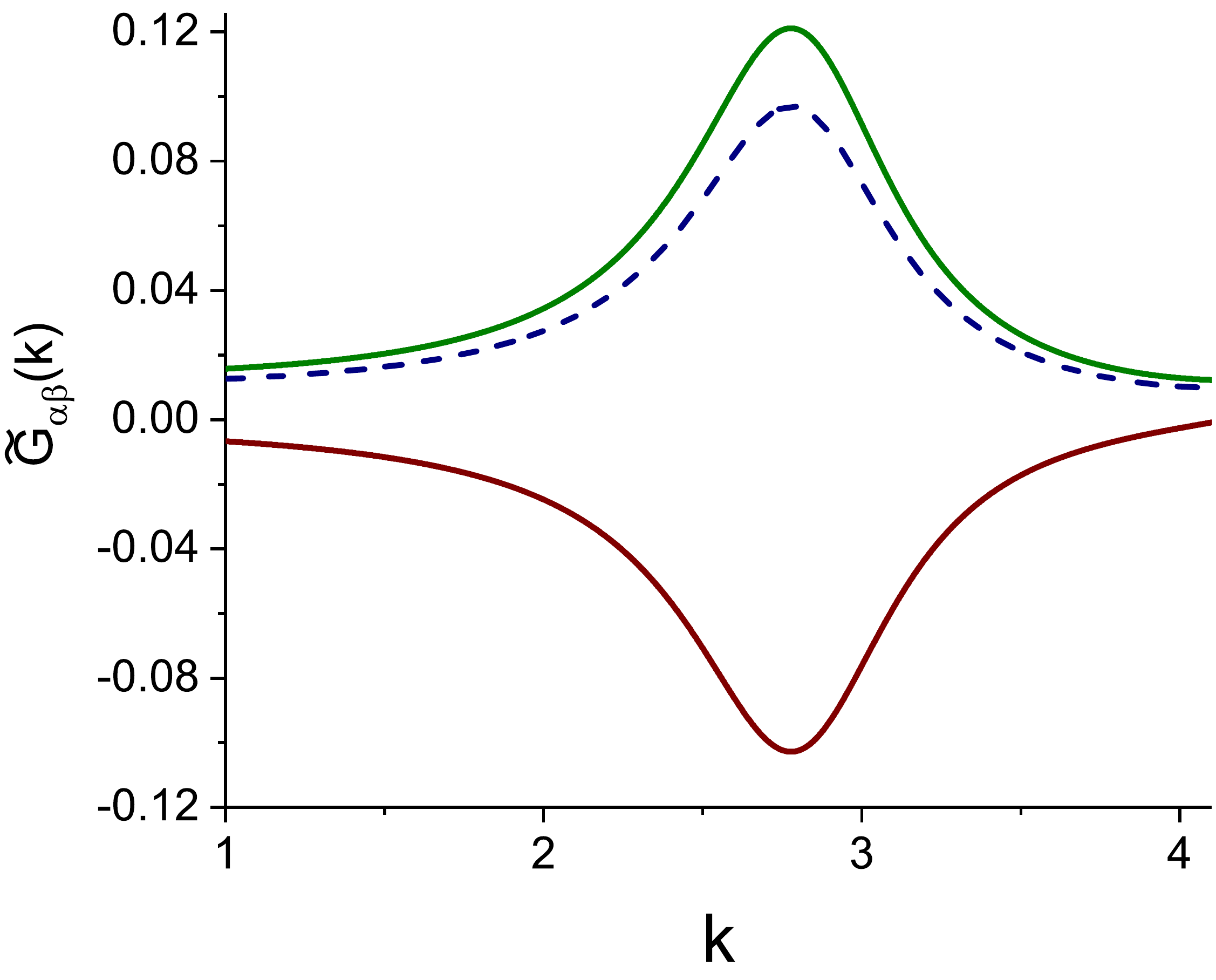} \qquad
	\includegraphics[clip,width=0.45\textwidth,angle=0]{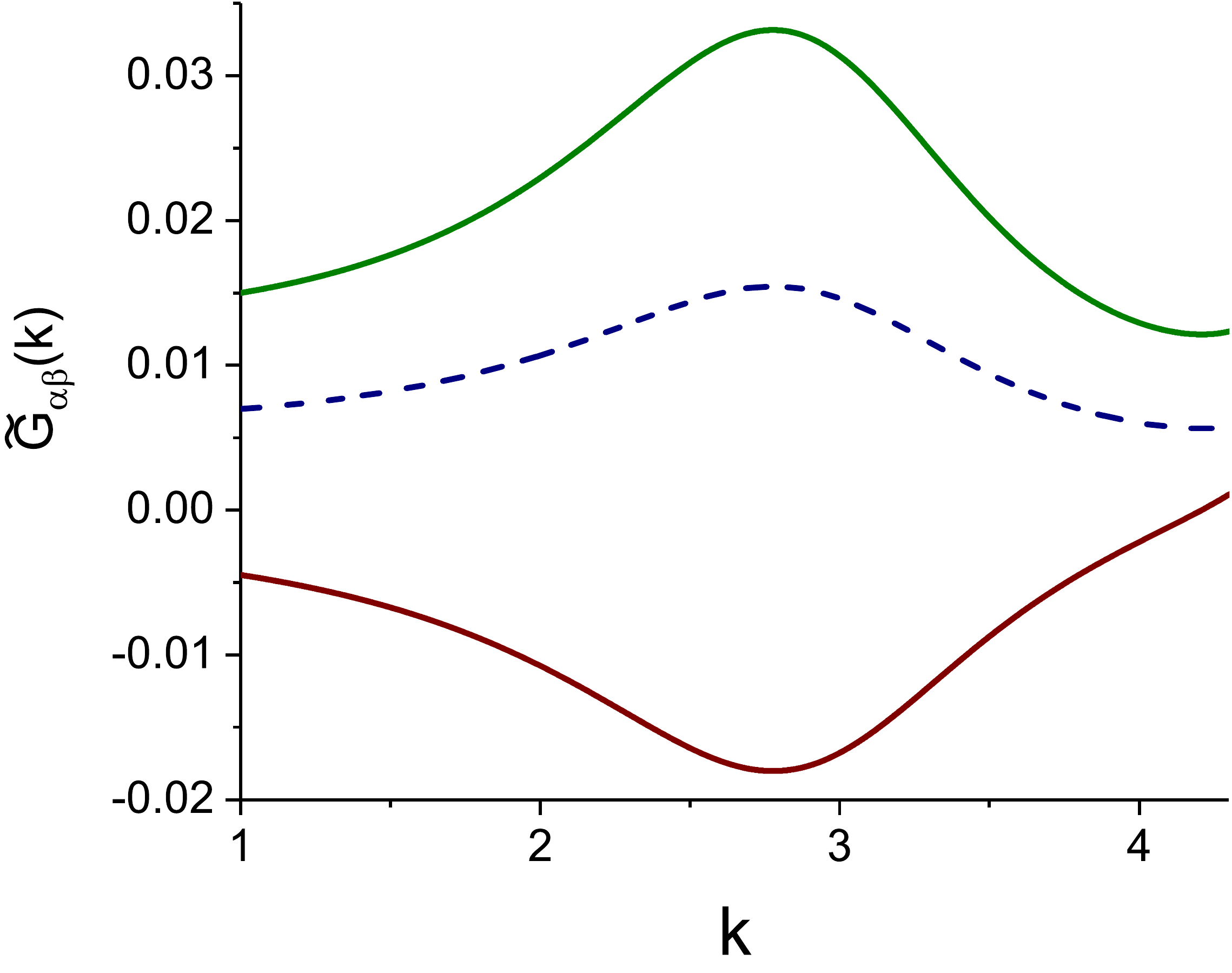}
	\caption{\label{G(k)-fl}
	(Colour online)	Correlation functions in Fourier representation  for $T^*=0.8$ and $\zeta=0.3$ 
		with the effect of fluctuations taken into account. Upper solid lines: $\tilde G_{11}(k)$, 
		dashed lines: $\tilde G_{22}(k)$ and lower solid lines: $\tilde G_{12}(k)$. Left-hand panel:
		$c=0.06$. Right-hand panel: $c=0.18$.
	}
\end{figure}

Using equations (\ref{C12_fl})--(\ref{G12-fl-sw}) and taking into account (\ref{C(k0)}),
we  calculate  the correlation functions in Fourier representation for $\zeta=0.3$ and for the temperature above and below  the $\lambda$-surface. In figure~\ref{GIJ-mf(k)} (right-hand panel), $\tilde G_{\alpha\beta}(k)$  are compared with the MF result for the same values of the temperature,  total volume fraction, and $c$. It is seen that the maxima (minimum) of the correlation functions for the temperature above the $\lambda$-surface become flat when the fluctuations are taken into account.   We  observe a nonmonotonous dependence of $\tilde G_{11}(k_{0})$  on $c$ with a  maximum at $c\approx 0.03$ (see figure~\ref{GIJ_c(k0)}, right-hand panel) and this behaviour is kept above and below the $\lambda$-surface. The correlation functions  $\tilde G_{22}(k_{0})$ and $\tilde G_{12}(k_{0})$ show a monotonous behaviour with an increasing $c$. In figure~\ref{G(k)-fl}, we present $\tilde G_{\alpha\beta}(k)$ for $T^*=0.8$ and $c=0.06$ and $c=0.18$ (thermodynamic states below  both the $\lambda$-surface and the gas-liquid spinodal). It is seen that the correlation functions have the same trend with an increase of $c$ as in the case of the `two mermaids and a peacock' model: all correlations become weaker, especially the correlations between the 
particles of a minority component. 

We should mention that the obtained extrema of the correlation functions in Fourier representation  
for $T^*=2.5$,  $\zeta=0.3$ and $c=0.06$ as well as for  $T^*=0.8$, $\zeta=0.3$ and  $c=0.18$, are flat. 
However, we assumed that high, narrow peaks are present. Since the theory should be self-consistent,
our findings signal that for these thermodynamic states our results cannot be accurate.
For  $T^*=0.8$, $\zeta=0.3$ and $c=0.06$, the extrema are more pronounced, and for this thermodynamic 
state better agreement with simulations can be expected.

\begin{figure}[!t]
	\centering
	\includegraphics[clip,width=0.5\textwidth,angle=0]{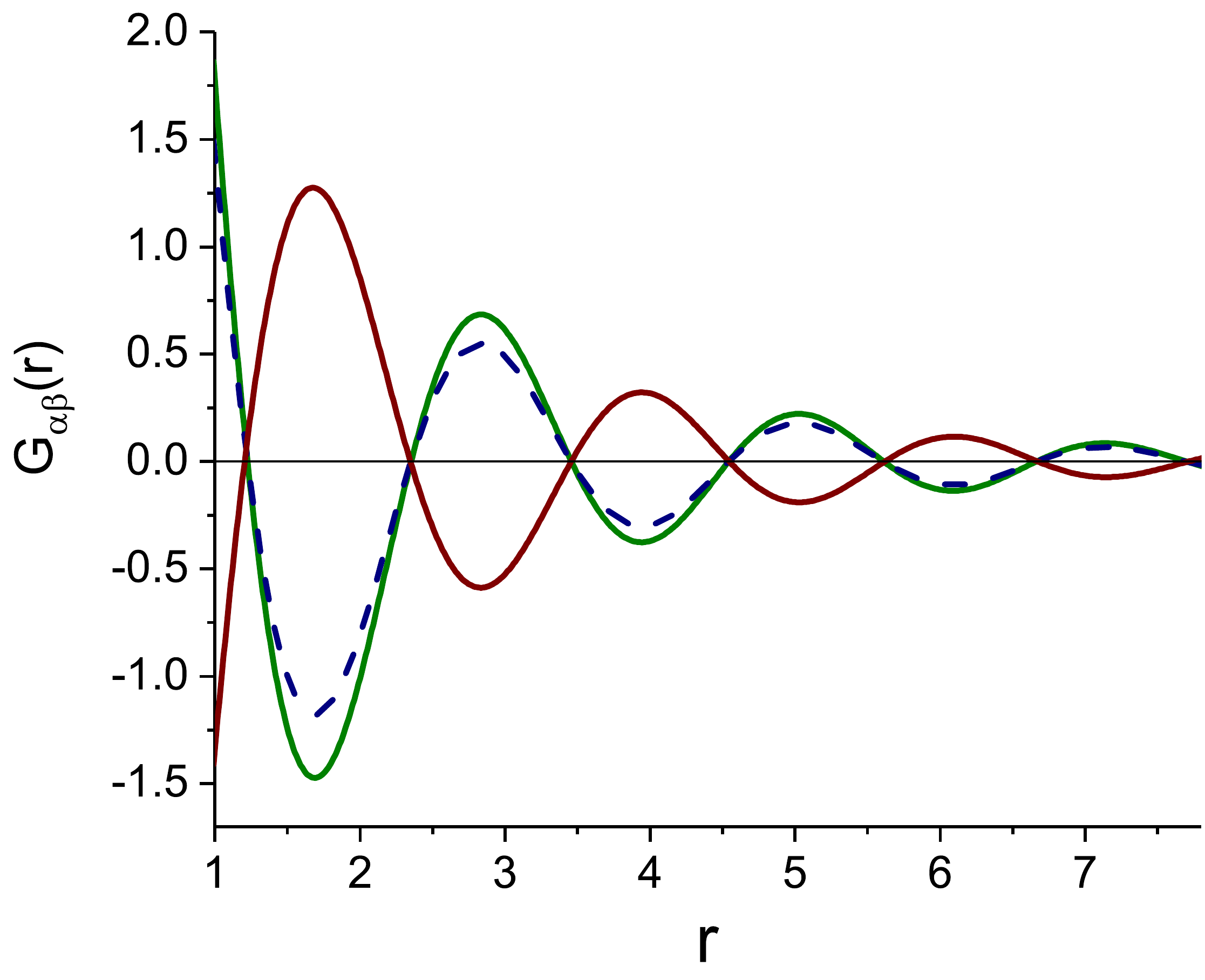}	
	\caption{\label{GIJ(r)_03}
		(Colour online)	Correlation functions in real space for $T^*=0.8$, $\zeta=0.3$ and $c=0.06$ with the effect of fluctuations taken into account.  $ G_{11}(r)$ (green upper solid line on the left), $ G_{22}(r)$ (dashed line) and $G_{12}(r)$ (red lower solid line on the left). 
	}
\end{figure}

In figure~\ref{GIJ(r)_03},  the correlation functions in real space are presented.  
As is seen, $G_{\alpha\beta}(r)$ shows the qualitative behaviour similar to the behaviour of the corresponding correlation functions of the `two mermaids and a peacock' model. The main difference is a period of oscillations which for this model is  $\approx 2\sigma$.

\vspace{-3mm}
\subsection{Simulations}
A binary mixture of particles interacting with the potentials~(\ref{V11-V22})--(\ref{V_sq-well}) was simulated using Monte Carlo technique in the $NVT$ ensemble. Particles ($N_{1}+N_{2}=15470$) are placed in a cubic box of the edge length $30\sigma$ with periodical boundary conditions in three directions. The volume fraction is $\zeta=0.3$.  Particles of both species are of the same diameter ($\sigma_{1}=\sigma_{2}=\sigma=1.0$). A cut-off radius of $2\sigma$ is used, which is the interaction range of the square well potential. Each system has run $10^{6}$ Monte Carlo steps for equilibration and  $10^{5}$ for production.

\begin{figure}[!b]
	\centering
	\includegraphics[clip,width=0.44\textwidth,angle=0]{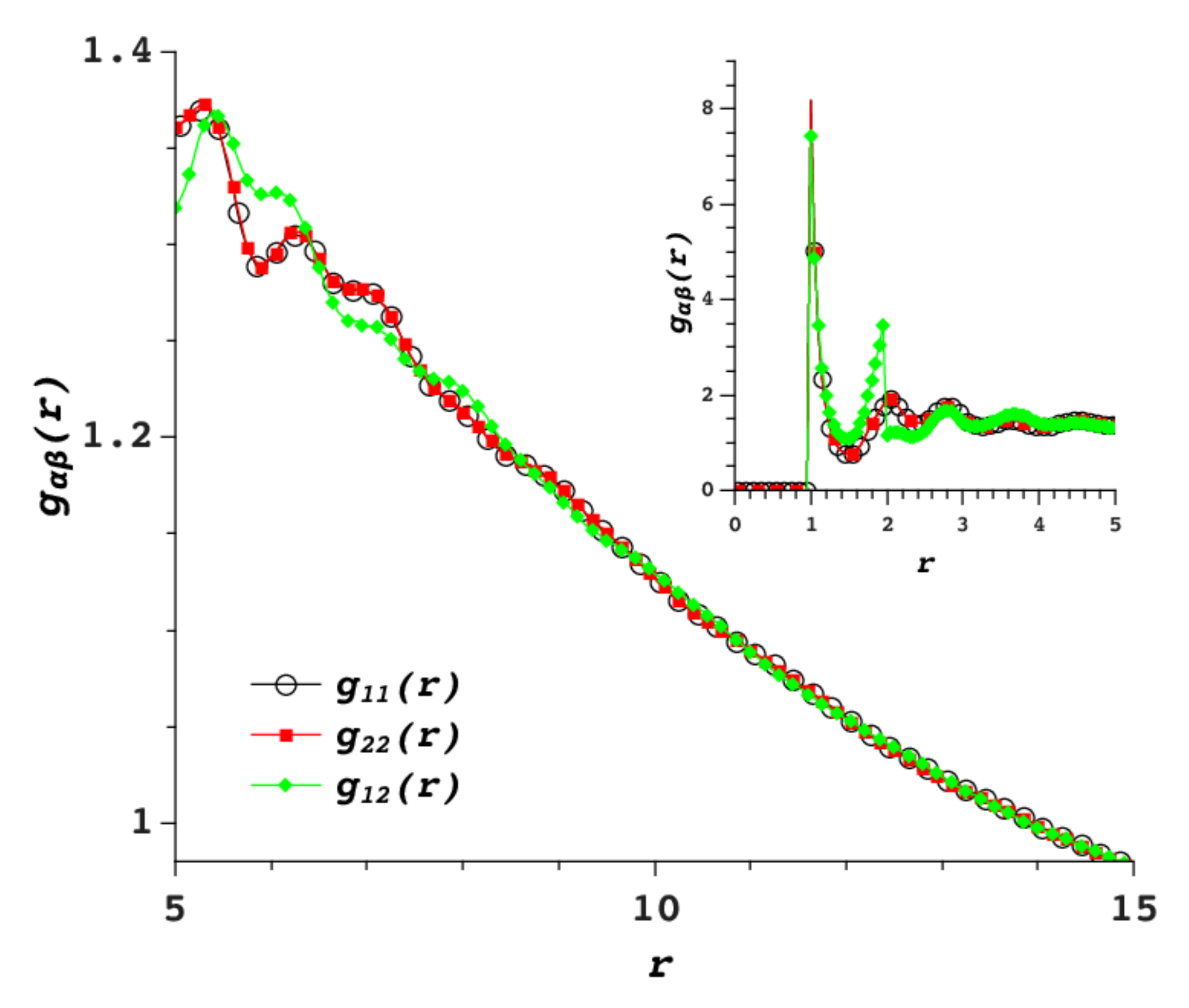}
	\includegraphics[clip,width=0.44\textwidth,angle=0]{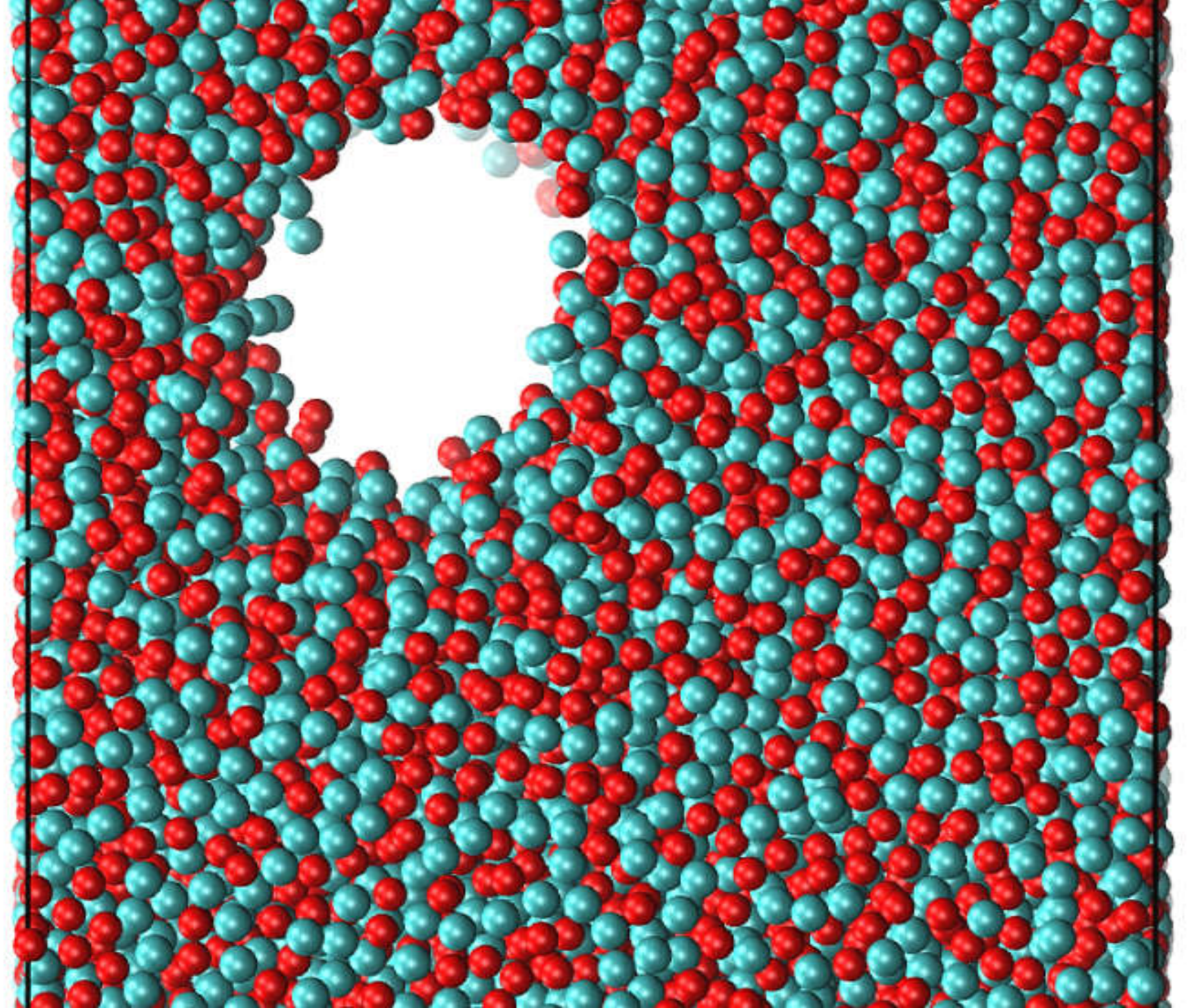}
	\caption{(Colour online) Pair distribution functions and a representative configuration  for $T^*=0.8$, and volume fractions 
		$\zeta_1=\zeta_2=0.15$ ($c=0$). }
	\label{sim_sqw-1}
\end{figure}

\begin{figure}[!b]
	\centering
	\includegraphics[clip,width=0.44\textwidth,angle=0]{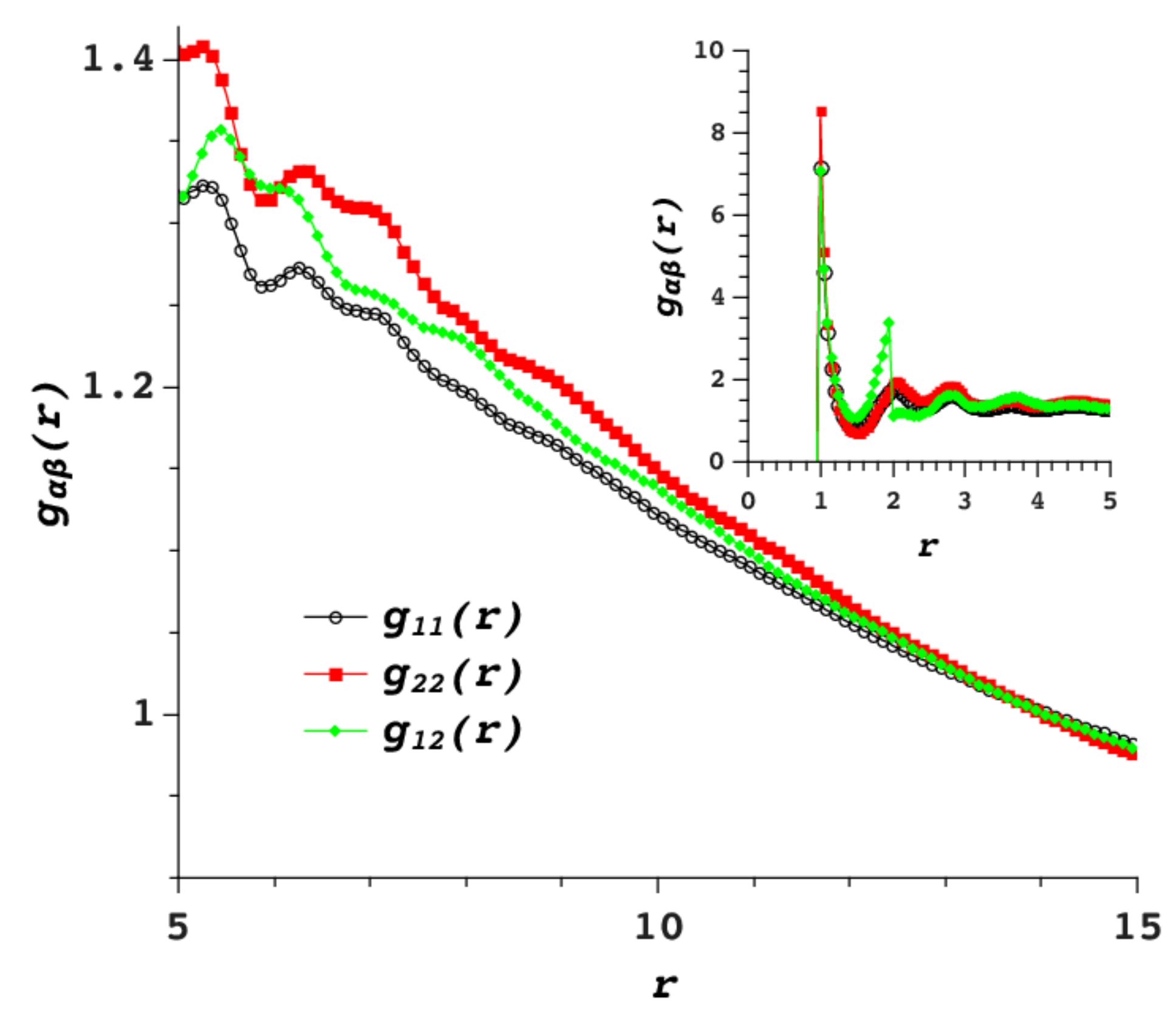}
	\includegraphics[clip,width=0.44\textwidth,angle=0]{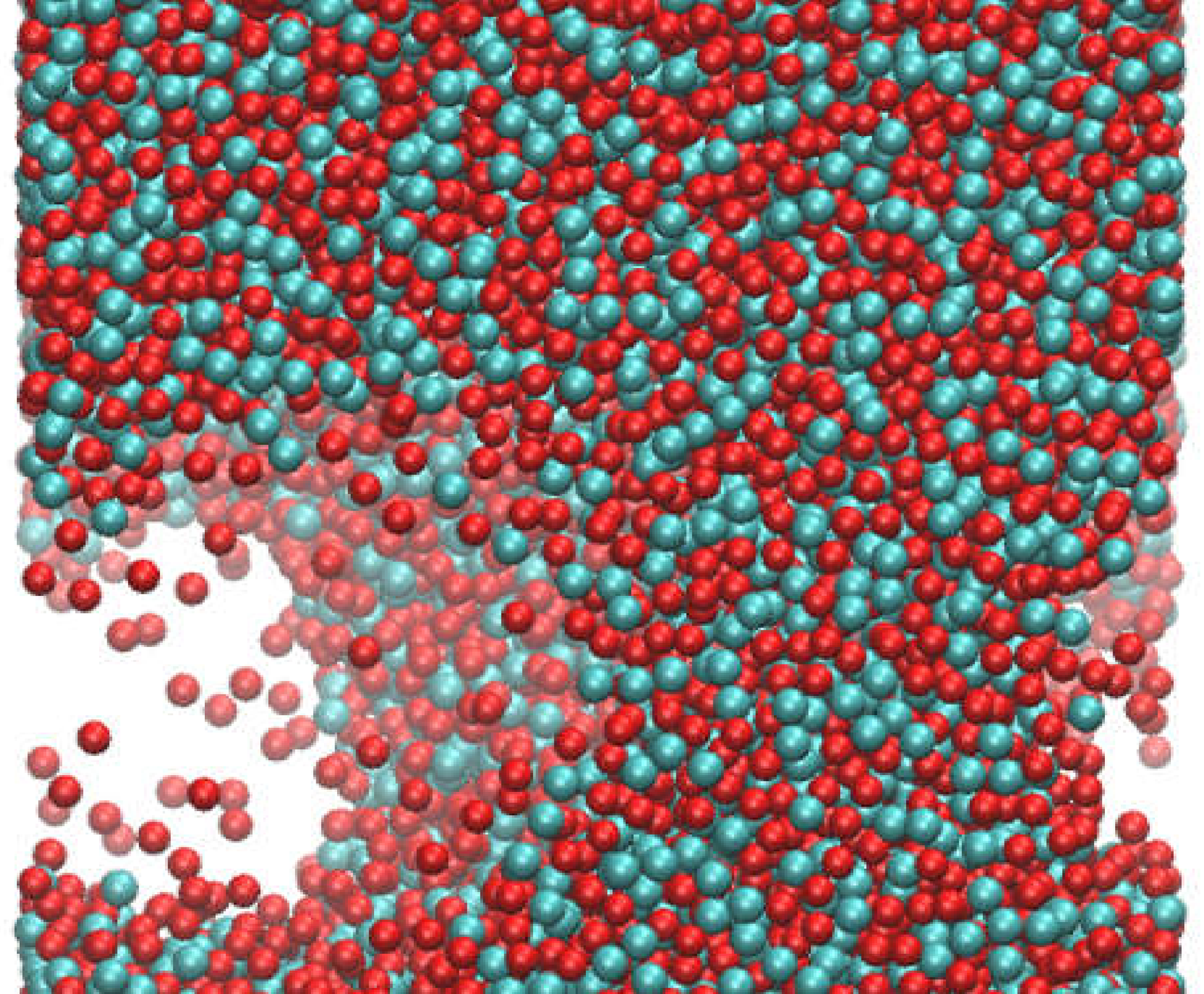}
	\caption{(Colour online) Pair distribution functions and a representative configuration  for $T^*=0.8$, and volume fractions 
		$\zeta_1=0.18$ and $\zeta_2=0.12$ ($c=0.06$). }
	\label{sim_sqw-2}
\end{figure}

\begin{figure}[!t]
	\centering
	\includegraphics[clip,width=0.44\textwidth,angle=0]{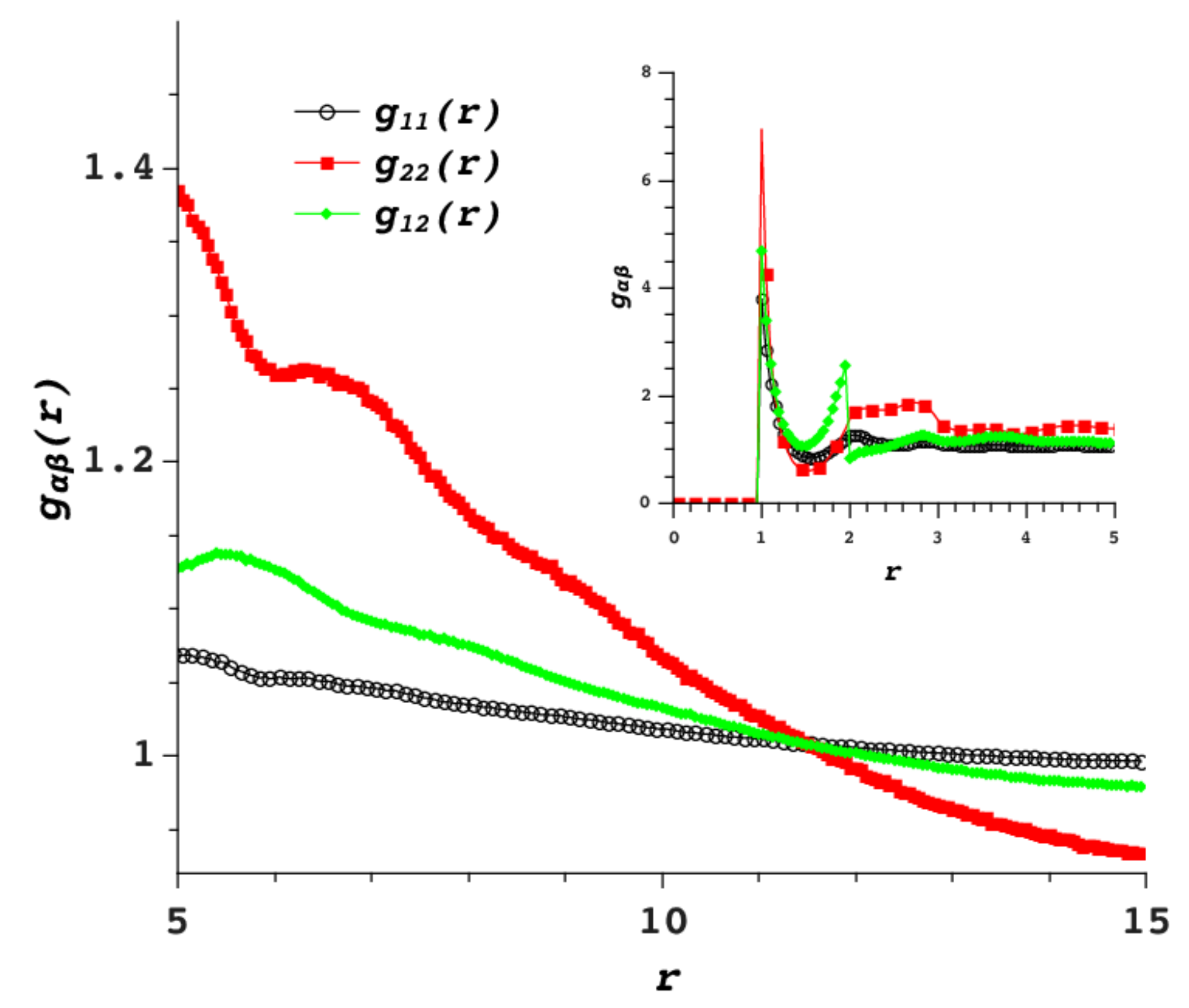}
	\includegraphics[clip,width=0.44\textwidth,angle=0]{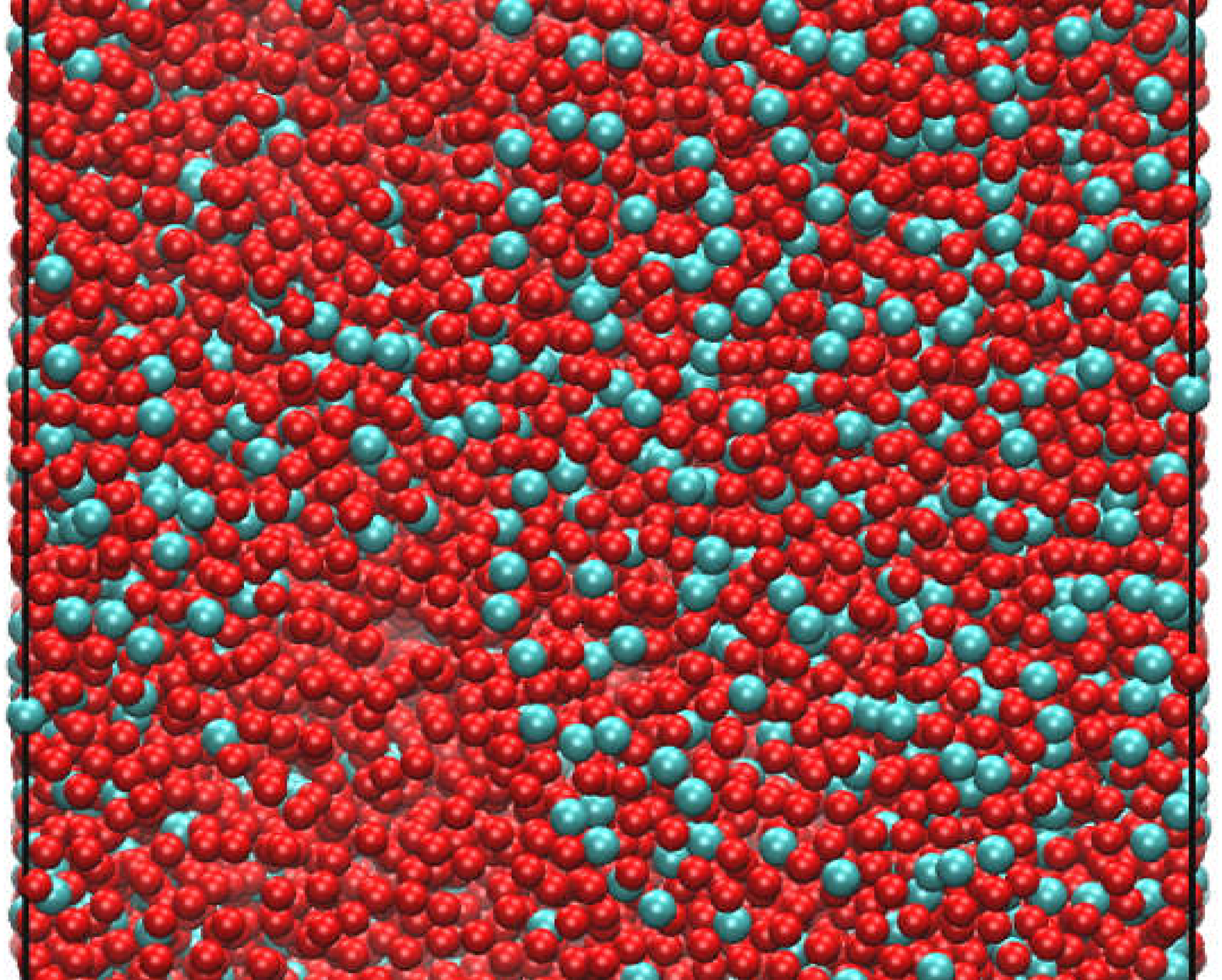}
	\caption{(Colour online) Pair distribution functions and a representative configuration  for $T^*=0.8$, and volume fractions 
		$\zeta_1=0.24$ and $\zeta_2=0.06$ ($c=0.18$). }
	\label{sim_sqw-3}
\end{figure}

In figures~\ref{sim_sqw-1}--\ref{sim_sqw-3}, the pair distribution functions and representative configurations for $T^*=0.8$, $\zeta=0.3$ and for three values of $c$ are presented.  The visualization suggests  that for $c=0$ and $0.06$ the dilute gas phase coexists with the dense inhomogeneous liquid (figures~\ref{sim_sqw-1}--\ref{sim_sqw-2}). 
For the strongly asymmetric case ($c=0.18$), the dense inhomogeneous liquid with approximately equal densities of the two components coexists with the one-component, very dense gas  (figure~\ref{sim_sqw-3}). In all cases, the  inhomogeneous liquid phase  mainly consists of stripes where the neighboring stripes are formed by the  particles of different type. However, this picture is less pronounced for $c = 0.18$.
\begin{figure}[!t]
	\centering
	\includegraphics[clip,width=0.42\textwidth,angle=0]{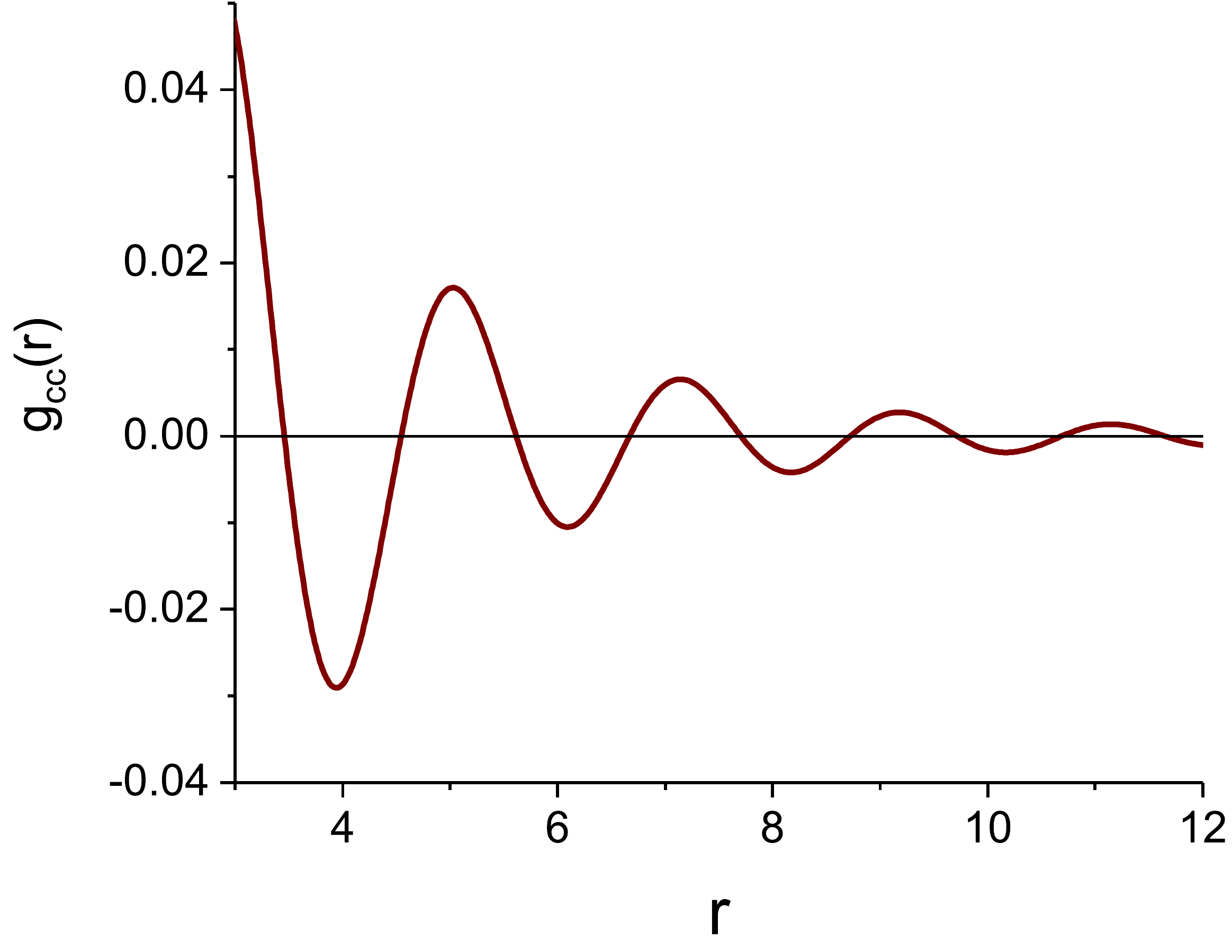} \qquad
	\includegraphics[clip,width=0.42\textwidth,angle=0]{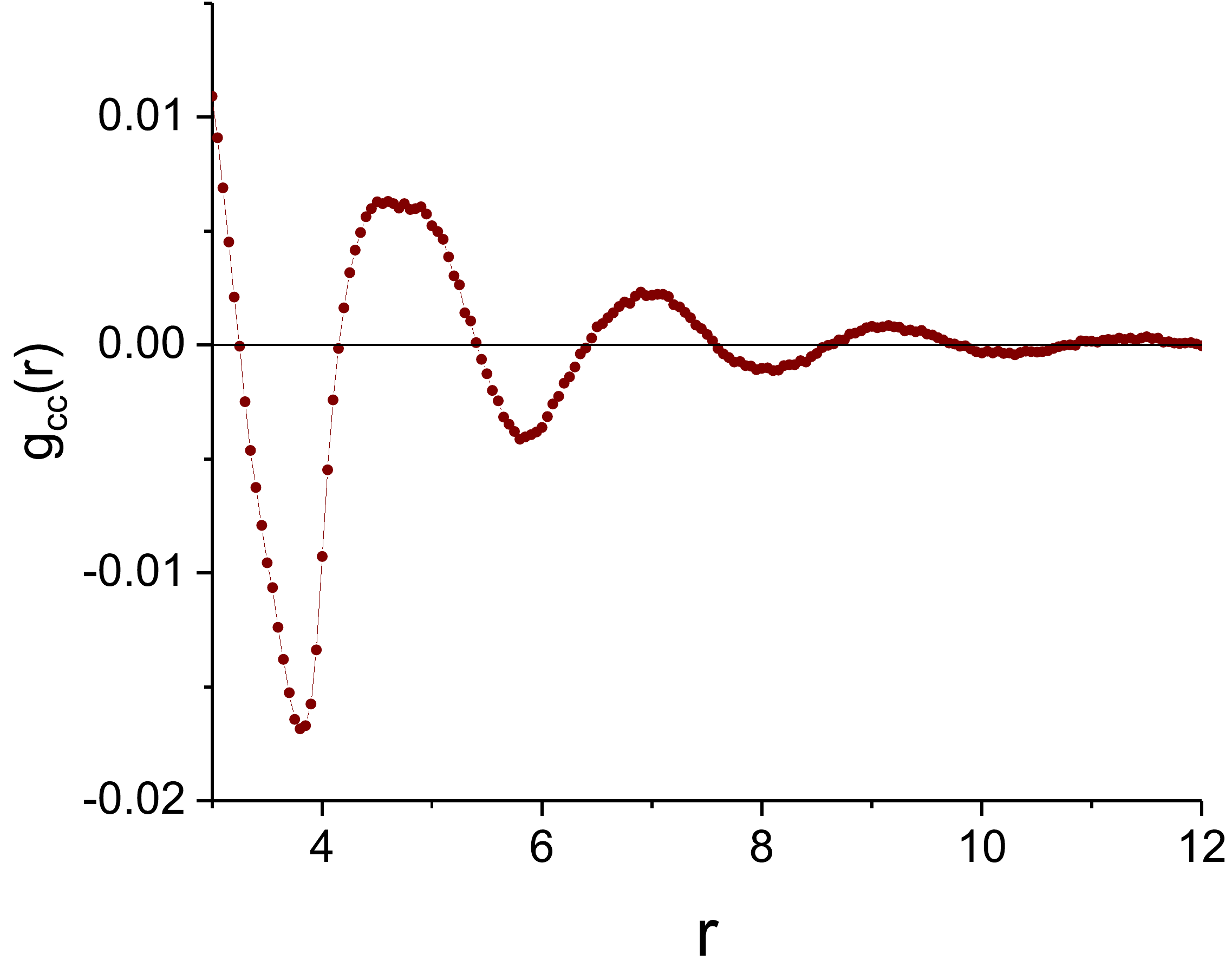}
	\caption{\label{g_cc_r} (Colour online)  Concentration-concentration distribution function at $T^*=0.8$,
		$\zeta=0.3$ and  $c=0.06$. Left-hand panel: theoretical results, right-hand panel: results of simulations. 
	}
\end{figure}
It should be noted that the structure of the disordered inhomogeneous phase at $\zeta=0.3$ is affected by the packing
of hard spheres. 
In order to separate the effect of hard-sphere packing  we calculate the concentration-concentration distribution 
function $g_{cc}(r)=c_{1}^{2}c_{2}^{2}(g_{11}(r)+g_{22}(r)-2g_{12}(r))$, where $c_{i}=\zeta_{i}/\zeta$ \cite{March1976}.  The results  are presented in figure~\ref{g_cc_r}.
Here, we compare  $g_{cc}(r)$  obtained  from the  theory (left-hand panel) and from the simulations (right-hand panel) 
for $T^*=0.8$, $\zeta=0.3$ and   $c=0.06$. One can see a rather good agreement between the theoretical
and simulation results:  the distribution functions $g_{cc}(r)$  show an oscillatory decay around zero with 
the period of damped oscillations $\lambda\approx 2$, and  the maxima and minima occur for very similar $r$  
in theory and simulations.  It is worth noting that in molten alkali halides, $g_{cc}(r)$ shows oscillations around zero, which extend over distances of at least 10~{\AA} \cite{March2002}.   Since the simulation results for $g_{cc}(r)$ are well reproduced by equation~(\ref{Gr}), the corresponding structure factor takes a maximum for $k=k_0$, and the peak shape agrees with our predictions.

\section{Conclusions} \label{sec5}
We have studied the effect of fluctuations on the correlation functions of a binary inhomogeneous mixture by using 
the mesoscopic density-functional theory. The theory is based on the Brazovskii-type approximation and allows one to 
take into account the fluctuation contribution. 
For a binary mixture in the disordered inhomogeneous phase, we have derived approximate equations
for the partial correlation functions in Fourier representation. Using  the above-mentioned approximations, we have calculated the correlation functions for
several state points above and below the MF instability for two particular models of a binary mixture of species A and B.
We have chosen models leading to a periodic structure on the length scale of the size of the particles,
and on the length scale 10 times larger. We did so in order to
compare the ordering and to verify the accuracy of the mesoscopic 
description for different length scales.

We have considered  the two models in the MF  
and when the fluctuations are taken into account. In the two models, we have limited  ourselves 
to an equal size of the hard cores of  the particles of both species. In the first mixture, the particles 
of the same type interact with the `mermaid' potential $V(r)$ (it has an attractive ‘head’
and a repulsive `tail') while A and B particles interact 
with a `peacock' potential $-V(r)$ which has an attractive `tail' and a repulsive `head'.
In the second mixture, there are only short-range  attractive interactions between  
species A and B beyond the hard core which is chosen in the square-well form. 
In both  models,  the attractions lead to a periodic arrangement of alternating species A and B.
However, the characteristic size of the regions rich in the particles of species A and B 
 differ significantly in the two models.  
In addition,  the gas-liquid phase separation can occur in the second model.

Despite different interaction potentials and different length scale of the local ordering, the correlation functions of the two models demonstrate similar properties. In order to verify the theoretical results and to visualize the  structure for different state points, we have performed MC simulations. It is shown that the theoretical predictions are consistent with the simulation results.

Our results can be valid only for the case of well-defined inhomogeneities. Well developed short-range order occurs
when the correlation functions in Fourier representation have high, narrow peaks at $k_0>0$. It is the case in the part of the phase diagram where 
the disordered phase looses stability in MF, but in reality it remains stable. Our theory agrees with simulations at a semiquantitative level for such state-points.

In the present paper, we have assumed equal size of the particles of different species. It is expected, however, that the periodic ordering should be enhanced when the size asymmetry increases. An increasing tendency for clustering with an increase of size asymmetry was observed in ionic systems in mesoscopic theory~\cite{ciach:11:2,ciach:07:0} and in simulations \cite{cheong:03:0,spohr:02:0}. Further work is necessary for the study of the correlation functions in  mixtures within the framework of the theory when the size asymmetry is taken into account.

\section*{Acknowledgements}
We would like to thank Vyacheslav Vikhrenko, Ruslan Lasovsky and Yaroslav Groda  for discussions 
and hospitality at the Belarusian State Technological University,
where a part of this work was done.
This project has received funding from the European Union Horizon 2020 research 
and innovation programme under the Marie
Sk\l{}odowska-Curie grant agreement No 734276 (CONIN).
An additional support in the years 2017--2020  has been granted  for the CONIN project by the Polish Ministry
of Science and Higher Education. 
Financial support from the National Science Center under grant No. 2015/19/B/ST3/03122 is also acknowledged.


\begin{thebibliography}{99}

\bibitem{blaaderen:05:0}
Leunissen M., Christova C., Hynninen A.P., Royal C., Campbell A., Imhof A.,
  Dijkstra M., van Roji R., van~Blaaderen A., Nature, 2005, \textbf{437}, 235,
  \doi{10.1038/nature03946}.

\bibitem{Bartlett2005}
Bartlett P., Campbell A.I., Phys. Rev. Lett., 2005, \textbf{95}, No.~12, 128302,
  \doi{10.1103/physrevlett.95.128302}.

\bibitem{Shimizu2015}
Shimizu K., Tariq M., Freitas A.A., P{\'{a}}dua A.A.H., Lopes J.N.C., J. Braz.
  Chem. Soc., 2015, \textbf{27}, 349--362, \doi{10.5935/0103-5053.20150274}.

\bibitem{Hayes2015}
Hayes R., Warr G.G., Atkin R., Chem. Rev., 2015, \textbf{115}, No.~13,
  6357--6426, \doi{10.1021/cr500411q}.

\bibitem{ciach:06:2}
Ciach A., Patsahan O., Phys. Rev. E, 2006, \textbf{74}, 021508,
  \doi{10.1103/PhysRevE.74.021508}.

\bibitem{patsahan:07:0}
Patsahan O., Ciach A., J. Phys.: Condens. Matter, 2007, \textbf{19}, 236203,
  \doi{10.1088/0953-8984/19/23/236203}.

\bibitem{stradner:04:0}
Stradner A., Sedgwick H., Cardinaux F., Poon W., Egelhaaf S., Schurtenberger
  P., Nature, 2004, \textbf{432}, 492, \doi{10.1038/nature03109}.

\bibitem{campbell:05:0}
Campbell A.I., Anderson V.J., van Duijneveldt J.S., Bartlett P., Phys. Rev.
  Lett., 2005, \textbf{94}, 208301, \doi{10.1103/PhysRevLett.94.208301}.

\bibitem{Sweatman:14:0}
Sweatman M.B., Fartaria R., Lue L., J. Chem. Phys., 2014, \textbf{140}, No.~12,
  124508, \doi{10.1063/1.4869109}.

\bibitem{candia:06:0}
De~Candia A., Gado E.D., Fierro A., Sator N., Tarzia M., Coniglio A., Phys.
  Rev. E, 2006, \textbf{74}, 010403(R), \doi{10.1103/PhysRevE.74.010403}.

\bibitem{santos:17:0}
Santos A.P., P\c{e}kalski J., Panagiotopoulos A.Z., Soft Matter, 2017,
  \textbf{13}, No.~44, 8055--8063,\\ \doi{10.1039/C7SM01721A}.

\bibitem{litniewski:19:0}
Litniewski M., Ciach A., J. Chem. Phys., 2019, \textbf{150}, 234702,
  \doi{10.1063/1.5102157}.

\bibitem{royall:18:0}
Royall C.P., Soft Matter, 2018, \textbf{14}, 4020, \doi{10.1039/c8sm00400e}.

\bibitem{bergman:19:0}
Bergman M., Garting T., Schurtenberger P., Stradner A., J. Phys. Chem. B, 2019,
  \textbf{123}, 2432,\\ \doi{10.1021/acs.jpcb.8b11781}.

\bibitem{falus:12:0}
Falus L.P., Fratini E., Chen W.R., Faraone A., Hong K., Baglioni P., Liu Y.,
  J. Phys.: Condens. Matter, 2012, \textbf{24}, 064114,
  \doi{10.1088/0953-8984/24/6/064114}.

\bibitem{ciach:08:1}
Ciach A., Phys. Rev. E, 2008, \textbf{78}, 061505,
  \doi{10.1103/PhysRevE.78.061505}.

\bibitem{ciach:13:0}
Ciach A., P{\k{e}k}alski J., G\'o\'zd\'z W.T., Soft Matter, 2013, \textbf{9},
  6301, \doi{10.1039/C3SM50668A}.

\bibitem{edelmann:16:0}
Edelmann M., Roth R., Phys. Rev. E, 2016, \textbf{93}, 062146,
  \doi{10.1103/PhysRevE.93.062146}.

\bibitem{pini:17:0}
Pini D., Parola A., Soft Matter, 2017, \textbf{13}, 9259,
  \doi{10.1039/C7SM02125A}.

\bibitem{zhuang:16:0}
Zhuang Y., Zhang K., Charbonneau P., Phys. Rev. Lett., 2016, \textbf{116},
  098301,\\ \doi{10.1103/PhysRevLett.116.098301}.

\bibitem{brazovskii:75:0}
Brazovskii S.A., Sov. Phys. JETP, 1975, \textbf{41}, 85.

\bibitem{leibler:80:0}
Leibler L., Macromolecules, 1980, \textbf{13}, 1602, \doi{10.1021/ma60078a047}.

\bibitem{podneks:96:0}
Podneks V.E., Hamley I.W., JETP Lett., 1996, \textbf{64}, 564,
  \doi{10.1134/1.567271}.

\bibitem{ciach:18:0}
Ciach A., Soft Matter, 2018, \textbf{14}, 5497, \doi{10.1039/C8SM00602D}.

\bibitem{ciach:11:2}
Ciach A., Mol. Phys., 2011, \textbf{109}, 1101,
  \doi{10.1080/00268976.2010.548343}.

\bibitem{ciach:18:1}
Ciach A., J. Mol. Liq., 2018, \textbf{270}, 138,
  \doi{10.1016/j.molliq.2017.10.002}.

\bibitem{ciach:12:0}
Ciach A., Patsahan O., Condens. Matter Phys., 2012, \textbf{15}, 23604,
  \doi{10.5488/CMP.15.23604}.

\bibitem{otero:18:0}
Otero-Mato J.M., Montes-Campos H., Cabeza O., Diddens D., Ciach A., Gallego
  L.J., Varela L.M., Phys. Chem. Chem. Phys., 2018, \textbf{20}, 30412,
  \doi{10.1039/C8CP05632C}.

\bibitem{ciach:16:0}
Ciach A., Adv. Biomembr. Lipid Self-Assembly, 2016, \textbf{23},
  61, \doi{10.1016/bs.abl.2015.12.004}.

\bibitem{ciach:10:1}
Ciach A., G\'o\'zd\'z W.T., Condens. Matter Phys., 2010, \textbf{13}, 23603, \doi{10.5488/CMP.13.23603}.

\bibitem{ciach:16:1}
Ciach A., Gozdz W.T., J. Phys.: Condens. Matter, 2016, \textbf{28}, 414010,
  \doi{10.1088/0953-8984/28/24/244004}.

\bibitem{March1976}
March N.H., Tosi M.P., Atomic Dynamics in Liquids, Macmillan Education, London, UK,
  1976,\\ \doi{10.1007/978-1-349-00929-9}.

\bibitem{March2002}
March N.H., Tosi M.P., Introduction to Liquid State Physics, World Scientific
  Publishing, Singapore, 2002, \doi{10.1142/4717}.

\bibitem{ciach:07:0}
Ciach A., G\'o\'zd\'z W.T., Stell G., Phys. Rev. E, 2007, \textbf{75}, 051505,
  \doi{10.1103/PhysRevE.75.051505}.

\bibitem{cheong:03:0}
Cheong D., Panagiotopoulos A., J. Chem. Phys., 2003, \textbf{119}, 8526,
  \doi{10.1063/1.1612473}.

\bibitem{spohr:02:0}
Spohr E., Hribar B., Vlachy V., J. Phys. Chem. B, 2002, \textbf{106}, 2343,
  \doi{10.1021/jp013811d}.

\end{thebibliography}

\ukrainianpart

\title{Вплив флуктуацій на кореляційні функції у неоднорідних сумішах}
\author{А. Цях\refaddr{label1}, О. Пацаган\refaddr{label2}, А. Мейра \refaddr{label3,label4} }
\addresses{
\addr{label1} Інститут фізичної хімії, Польська академія наук, м. Варшава, вул. Каспшака, 44/32, Польща
\addr{label2} Інститут фізики конденсованих систем Національної академії наук України, \\вул. Свєнціцького, 1, 79011 Львів, Україна
\addr{label3} IFLYSIB (UNLP, CONICET), 59 No. 789, B1900BTE м. Ла-Плата, Аргентина
\addr{label4}Національний технологічний університет - регіональний факультет Ла-Плата, кафедра машинобудування,
м. Ла-Плата, Аргентина
}
%
%
%

\makeukrtitle 

\begin{abstract}
\tolerance=3000%
В рамках мезоскопічної теорії [A. Ciach, Mol. Phys., 2011, {\bf 109}, 1101] виведено наближені вирази для кореляційних функцій в бінарних неоднорідних сумішах. Флуктуаційний внесок враховується в наближенні Бразовського. Отримано явні результати для двох модельних систем. В обох моделях, діаметри твердого кору частинок є однаковими, а взаємодії  сприяють періодичному чергуванню в розташуванні  сортів A і B. Проте, оптимальна відстань між сортами  A і B суттєво різниться в двох моделях. Теоретичні результати для різних значень температури і об'ємних фракцій двох компонент порівнюються  з результатами моделювання методом Монте Карло, а структури проілюстровані миттєвими знимками системи. Незважаючи на те, що потенціали взаємодії і масштаби локального впорядкування в двох моделях є різними,  властивості  кореляційних функцій цих моделей є дуже подібними. 
\keywords кореляційні функції, неоднорідні суміші, мезоскопічна теорія
\end{abstract}

\end{document}